\documentclass[sigconf]{acmart}

\AtBeginDocument{%
  \providecommand\BibTeX{{%
    \normalfont B\kern-0.5em{\scshape i\kern-0.25em b}\kern-0.8em\TeX}}}

\copyrightyear{2023}
\acmYear{2023}
\setcopyright{acmlicensed}\acmConference[FAccT '23]{2023 ACM Conference on Fairness, Accountability, and Transparency}{June 12--15, 2023}{Chicago, IL, USA}
\acmBooktitle{2023 ACM Conference on Fairness, Accountability, and Transparency (FAccT '23), June 12--15, 2023, Chicago, IL, USA}
\acmPrice{15.00}
\acmDOI{10.1145/3593013.3594108}
\acmISBN{979-8-4007-0192-4/23/06}

\usepackage{subcaption}
\usepackage{csquotes}
\usepackage{xspace}
\usepackage{array}

\newcommand{\etal}{\emph{et al.}\@\xspace}

\newcommand{\eg}{\emph{e.g.}\xspace}

\newcommand{\fairfuse}{FairFuse\xspace}
\newcommand{\confuse}{ConsensusFuse\xspace}

\newcommand{\hs}[1]{{#1}}

\begin{document}




\title[Help or Hinder? Fairness in Visualizations for Consensus Ranking]{Help or Hinder? Evaluating the Impact of Fairness Metrics and Algorithms in Visualizations for Consensus Ranking}

\author{Hilson Shrestha}
\affiliation{%
  \institution{Worcester Polytechnic Institute}
  \city{Worcester}
  \state{Massachussetts}
  \country{USA}
}
\email{hshrestha@wpi.edu}

\author{Kathleen Cachel}
\affiliation{%
  \institution{Worcester Polytechnic Institute}
  \city{Worcester}
  \state{Massachussetts}
  \country{USA}
}
\email{kcachel@wpi.edu}

\author{Mallak  Alkhathlan}
\affiliation{%
  \institution{Worcester Polytechnic Institute}
  \city{Worcester}
  \state{Massachussetts}
  \country{USA}
}
\email{malkhatlan@wpi.edu}

\author{Elke Rundensteiner}
\affiliation{%
  \institution{Worcester Polytechnic Institute}
  \city{Worcester}
  \state{Massachussetts}
  \country{USA}
}
\email{rundenst@wpi.edu}

\author{Lane Harrison}
\affiliation{%
  \institution{Worcester Polytechnic Institute}
  \city{Worcester}
  \state{Massachussetts}
  \country{USA}
}
\email{ltharrison@wpi.edu}

\renewcommand{\shortauthors}{Shrestha, et al.}

\begin{abstract}
For applications where multiple stakeholders provide recommendations, a fair consensus ranking must not only ensure that the preferences of rankers are well represented, but must also mitigate disadvantages among socio-demographic groups in the final result. 
However, there is little empirical guidance on the value or challenges of visualizing and integrating fairness metrics and algorithms into human-in-the-loop systems to aid decision-makers.
In this work, we design a study to 
analyze the effectiveness of integrating such fairness metrics-based visualization and algorithms.
We explore this through a task-based crowdsourced experiment comparing an interactive visualization system for constructing consensus rankings, \confuse, with a similar system that includes visual encodings of fairness metrics and fair-rank generation algorithms, \fairfuse. We analyze the measure of fairness, agreement of rankers' decisions, and user interactions in constructing the fair consensus ranking across
these two systems.
In our study with 200 participants, results suggest that providing these fairness-oriented support features 
nudges users to align their decision with the fairness metrics while minimizing the tedious process of manually having to amend the consensus ranking.
We discuss the implications of these results for the design of next-generation fairness oriented-systems and along with 
emerging directions for future research.
\end{abstract}




\maketitle


\section{Introduction}
\begin{figure*}
  \includegraphics[width=\textwidth]{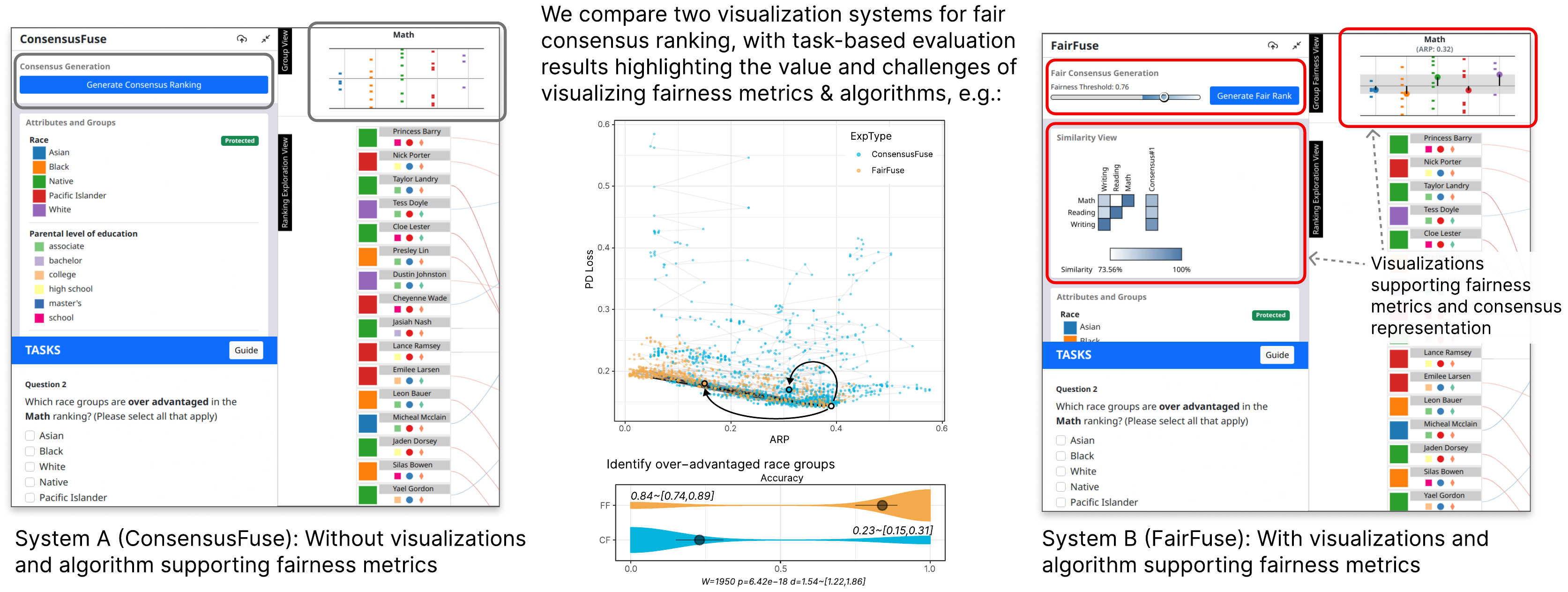}
  \caption{Can visualization-enabled fairness metrics aid fairness related tasks? We compare two systems: A: \confuse, a visualization that enables fairness comparison only by interactive visual displays of underlying items. B: \fairfuse, a similar system which visualizes additional fairness metrics and provides a fair-rank generation algorithm. We find positive impact of embedding fairness metrics and algorithms into visualization supporting consensus ranking scenarios, but also certain risks.}
  \label{fig:teaser}
\end{figure*}

The broader fairness community has developed a vast array of metrics and algorithms that conceptualize, measure, and systematize definitions of fairness, in part to guide decision-making in computing contexts.
One popular medium for operationalizing these metrics in user-centric computing systems are interactive visualizations. Such visualizations  can provide increased transparency across  the underlying data,
the decision algorithms applied to the data, and the corresponding fairness properties expressed by fairness metrics, among other benefits.

Several recent efforts highlight the inherent promise of interactive visualization for advancing goals in the fairness community, such as work from Mitchell \etal and Crisan \etal on model cards \cite{mitchell2019model, crisan2022interactive}, and Van Berkel \etal on examining the value of visualization over text for communicating fairness concepts \cite{van2021effect}.
However, efforts combining visualization approaches and fairness metrics and algorithms  raise 
both 
challenges as well as unique 
opportunities in this space.
Can visualizations aid some fairness-related tasks, but hinder others? 
Should fairness metrics be visualized by tightly integrating them  with the underlying data items, or separately through popular visualization techniques such as coordinated multiple views? 
Might some visualizations even mislead or otherwise reduce the agency of users in achieving fairness in decision-making contexts?

In this paper, we explore these broad questions through a particular instance of a controlled task-based visualization study. 
Our context is fair ranking problems,
where prior works 
have focused on achieving group fairness, i.e. treating all groups in the ranking similarly, \eg \cite{kuhlman2020rank, cachel2022manirank, shrestha2022fairfuse, ahn2019fairsight, xie2021fairrankvis}.
We begin with a recently 
published fairness visualization with an available open source system, \fairfuse \cite{shrestha2022fairfuse}, 
which visualizes both the candidate items to be ranked and various fairness metrics  such as Favored Pair Representation (FPR), Attribute Rank Parity (ARP), and  PD Loss \cite{cachel2022manirank}.
We adapt \fairfuse into a new system, \confuse, by removing visualized fairness metrics and algorithms (Figure \ref{fig:fairfuse-system}). 
We conduct a goal and activity analysis (\eg \cite{hindalong2020towards, hindalong2022abstractions, munzner2009nested} to define fairness-oriented tasks in ranking contexts (Table \ref{table:goals-analysis}, \ref{table:tasks-analysis}).
We distill these goals and activities into a set of evaluation tasks, with measurable outcomes (Table \ref{tab:questions-participants}).


\hs{With the two systems and the above-identified tasks in place, we conduct a controlled experiment with $n=200$ participants (100 per condition).}
Results generally validate that visualizing fairness metrics leads to notably increased accuracy in key fairness-related tasks.
However, deeper analysis of measures, exploration behavior, and participant explanations reveal nuance, challenges and risks in visualizing fairness metrics.
We discuss findings, such as for instance the fact that the presence of algorithm-driven fairness schemes tended to ``shift'' participants' exploration and ultimate decisions in a ranking task. 
We also develop a set of takeaways highlighting where visualized fairness generally tends to help, but also where it may hinder users in decision-making contexts.

{\bf Contributions.}
Taken together, our work makes the following contributions: 

\begin{itemize}
    \item A task-based evaluation comparing a system that visualizes fairness metrics/algorithm results against a control with equivalent functionality, sans metrics/algorithms.
    \item Results that generally validate the value of visualizing fairness metrics/algorithms for rank-focused contexts.
    \item Additional analyses that highlight particular challenges in visualization design for fairness, including risks and tensions in fairness interface design that may require substantial future effort to resolve.
\end{itemize}

\section{Related Work}


\hs{Research to date has developed several fairness-related visualization systems, such as tools for consensus-building and ranking-based tasks. 
However, throughout these efforts, there remains a lack of empirical evidence examining the value and challenges of visualizing and incorporating fairness metrics and algorithms into human-in-the-loop systems.
Here we review several of the systems, efforts, and concepts we draw from when designing the present experiment.}

\subsection{Visualizing and Presenting Fairness in Information Systems}

Much of the work in algorithmic fairness in recent years  has focused on proposing various conceptualizations
of fairness, along with algorithmic techniques for ensuring these definitions are met in decision-making processes.
Comparatively less work has proposed fairness-oriented visualization systems or studied the merits of visual
representations of fairness and bias in decision-making.

\subsubsection{Fairness Visualization Tools and Toolkits}

The design of interactive or visual systems has predominately focused on highlighting and providing recourse for socio-demographic bias in classification tasks \cite{bantilan2018themis, saleiro2018aequitas, bird2020fairlearn, xie2021fairrankvis, wexler2019if}.
The focus on classification-based machine learning models mirrors the attention of the larger algorithmic fairness community, namely, where ``Fair-ML" gained prominence in the context of binary classification. 
Many tools 
have been developed to detect algorithmic biases and to evaluate and compare different machine learning models concerning fairness \cite{tramer2017fairtest, bellamy2018ai, johnson2022fairkit}.
Crisan \etal and Mitchell \etal \cite{crisan2022interactive, mitchell2019model} proposed visual model cards for documenting models for better transparency.
Recent visualization research has focused on addressing
group bias discovery and the interpretation of intersectional bias \cite{cabrera2019fairvis, munechika2022visual}.
In the context of rankings, Yang \etal \cite{yang2018nutritional} provided 
``nutritional facts" for the fairness of rankings, Ahn \etal \cite{ahn2019fairsight} proposed an interactive system for building fair rankings, and Xie \etal \cite{xie2021fairrankvis} introduced a visual system for fairness comparing rankings produced from graph mining recommender algorithms.

\subsubsection{Evaluation of Fairness-Oriented Toolkits}

Several researchers assessed toolkits that incorporate fairness into their process.
Mashhadi \etal \cite{mashhadi2022case} studied the impact of the visualization styles of six open-source fair classification toolkits on student learning of fairness criteria.
Lee \etal \cite{lee2021landscape} evaluated the capabilities of open-source fairness toolkits and their suitability for commercial use through practitioner interviews and surveys. 
They found that many toolkits that contained visual representations of fairness were difficult for non-technical users to understand, even in tools like the What-If Tool \cite{wexler2019if}, which were designed for broader audiences. 
Richardson \etal \cite{richardson2021towards} conducted interviews with machine learning practitioners to create a rubric for evaluating fairness toolkits.
While there has been a surge in the development of fairness toolkits, Deng \etal \cite {deng2022exploring} have highlighted gaps between fairness toolkits' capabilities and practitioners' needs.

\subsubsection{Evaluation on Presentation of Fairness Information}
Studies have evaluated the
presentation of  fairness related information in different scenarios.
Van Berkel \etal \cite{van2021effect} compared the 
perceived fairness level between text and scatterplot visualization techniques. The study found that the
scatterplot visualization technique resulted in a lower fairness perception than 
text.
Saxena \etal \cite{saxena2019fairness} investigated people's attitudes towards algorithmic definitions of fairness and found that people considered calibrated models, such as ratios, fairer than equal or meritocratic distributions in the context of loan decisions.
Similar studies found that people perceive demographic parity and equalized odds as fair, depending on the scenario. 
Cheng \etal \cite{cheng2021soliciting} compared three group fairness approaches in a child maltreatment predictive system.
They  found that people mostly supported equalized odds, followed by statistical parity and unawareness. 
Srivastava \etal \cite{srivastava2019mathematical} found that people prefer demographic parity among the 6 different notions of group fairness.
Harrison \etal \cite{harrison2020empirical} conducted a user study on the perceived fairness of machine learning models in the criminal justice context and found conflicts between various inconsistent definitions of fairness.
Nevertheless, Hannan \etal \cite{hannan2021gets} showed that the factors of "what" and "who" matter in fairness perceptions and that the context of algorithmic fairness is more important in some domains than others.

\subsection{Tools and Evaluation Studies on Consensus Building}

Visualization systems have been designed to aid decision-makers in inspecting multiple stakeholders' preferences to reach a consensus decision \cite{bajracharya2018interactive, carenini2004valuecharts, dimara2017dcpairs, hansen2008new, hayez2012d, hong2018collaborative,liu2018consensus, pham2016qstack, weng2018srvis, weng2018homefinder, shah2014collaborative, mustajoki2000web}.
A subset of these tools consider the setting, like ours, in which stakeholder preferences are encoded as rankings \cite{hindalong2020towards, carenini2004valuecharts, liu2018consensus}. 
Liu \etal \cite{liu2018consensus} evaluated a between-subjects experiment to assess the effectiveness of their proposed tool, ConsensUs, designed for multiple stakeholders to rate and select candidates.
They found that visualizations helped surface stakeholder disagreement that otherwise would have gone undetected. 
Hindalong \etal \cite{hindalong2022abstractions} perform an evaluation 
study of six tools (both visualization-focused systems and commercial systems that implicitly allow for stakeholder preference inspection), including the systems of \cite{hindalong2020towards, liu2018consensus, carenini2004valuecharts}. 
The corresponding evaluation studies are focused on how well these tools help achieve consensus outcomes -- yet
none consider the employment of consensus generation algorithms \cite{copeland1951reasonable, kemeny1959mathematics, schulze2018schulze, borda1781mathematical}. In contrast, we study consensus building when decision makers are supported by fair consensus rank generation algorithms and when
 fairness metrics are presented visually throughout the process.

\subsection{Tools for Ranking-based Tasks and Corresponding Evaluation Studies}
Interactive systems and evaluation studies of visualization paradigms have been developed specifically for ranking data. Gratzl \etal \cite{gratzl2013lineup} propose a visualization system, LineUp, to compare ranked items along multiple attributes. Their qualitative evaluation study found that visualizations helped people perform challenging ranking-based tasks faster. Wall \etal \cite{wall2017podium} presented Podium, a visual analytics tool for helping users define a ranking function combining multiple criteria according to their interactions with a subset of the ranked data. Behrisch \etal \cite{behrisch2013visual} presented a visual system to compare similarities and differences of pairs of rankings using small multiple views of glyphs. However, while the above works target rank-oriented workflows, they neither consider the problem of visually comparing a consensus ranking vis-a-vis the stakeholder's respective base rankings nor how fairness metrics should be incorporated visually throughout the consensus ranking process.

\section{Visualization and Interaction Design}


\begin{figure*}[tb]
  \centering
  \includegraphics[width=\linewidth]{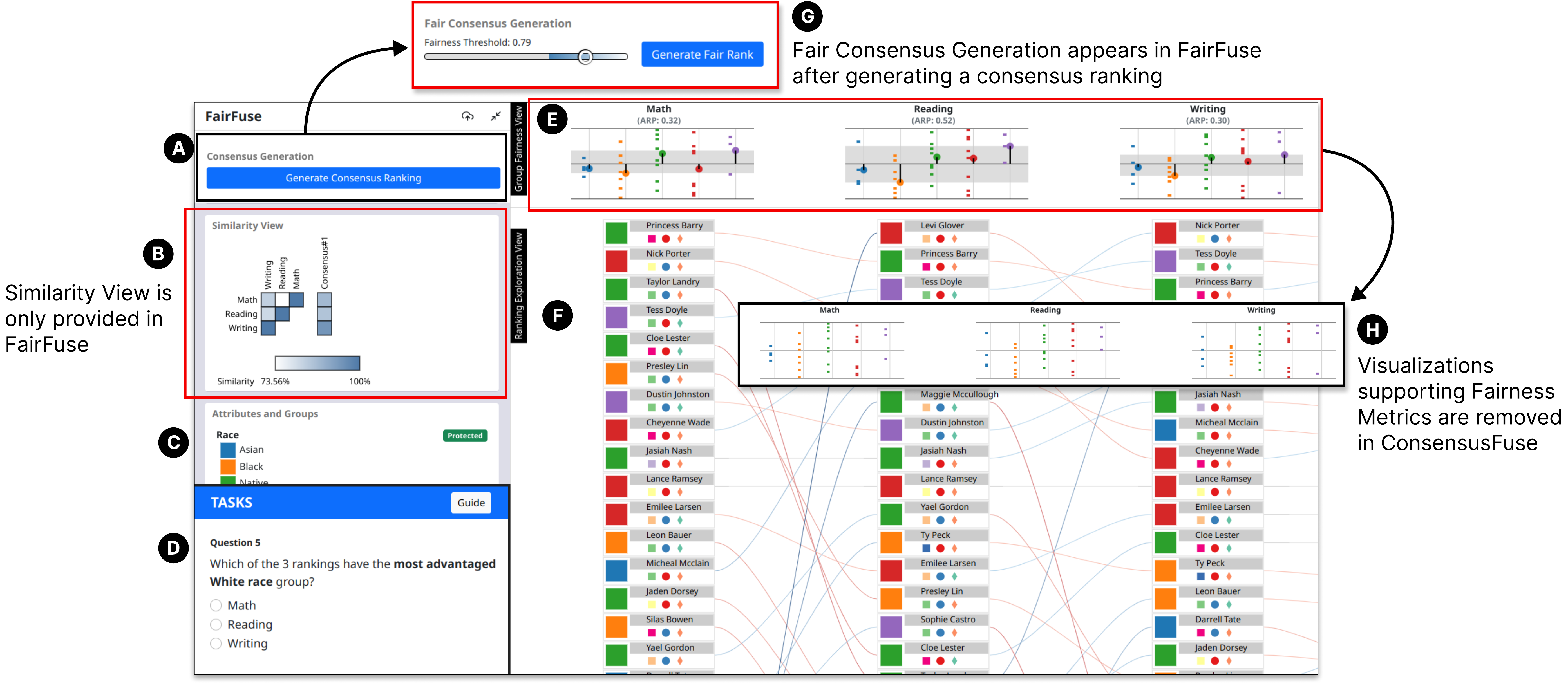}
  \caption{\fairfuse and \confuse System Designs with changes in visualizations related to fair consensus generation. A) Consensus Generation, B) Similarity View (in \fairfuse), C) Attributes Legend, D) Tasks presented to the participants, E) Group Fairness View (in \fairfuse), F) Ranking Exploration View, G) Fair Consensus Generation (in \fairfuse), H) Group View (in \confuse).
  }
  \label{fig:fairfuse-system}
\end{figure*}

To be able to study the challenges and opportunities in visualizing fairness metrics and algorithms,
we have  modified the \fairfuse system \cite{shrestha2022fairfuse}
to create two system variations.
The \fairfuse system employed task abstraction methodologies following procedures from Lam \etal \cite{lam2017bridging} and recent works on group decision-making by Hindalong \etal \cite{hindalong2020towards, hindalong2022abstractions}. Table \ref{table:goals-analysis} outlines the goals and sub-goals for generating and analyzing fair consensus rankings 
that combines the preferences of multiple rankers (base rankings) into a single consensus ranking. For each sub-goal, we identified a set of visualization activities (Table \ref{table:tasks-analysis}) based on a  widely used method in the visualization literature 
\cite{brehmer2013multi}, leading to the design and implementation of several views.

\begin{table}[tb]
  \caption{Generic goals for rankings inspection and fair ranking generation and analysis}
  \label{table:goals-analysis}
  \centering
  \begin{tabular}{llp{.8\linewidth}}
    \toprule
    \multicolumn{3}{l}{\textbf{GENERIC GOAL}} \\
    \midrule
    
    \multicolumn{1}{l}{G1} & \multicolumn{2}{l}{\textbf{Characterize Differences in Base Rankings}}\\
    \multicolumn{1}{l}{}   & \multicolumn{1}{l}{a} & Discover (dis)agreement on each candidate between rankings \\
    \multicolumn{1}{l}{}   & \multicolumn{1}{l}{b} & Assess the discrepancy of candidates' position between base rankings \\
    \midrule
    \multicolumn{1}{l}{G2} & \multicolumn{2}{l}{\textbf{Investigate Protected Attribute}} \\
    \multicolumn{1}{l}{}   & \multicolumn{1}{l}{a} & Discover protected attribute groups of the candidates \\
    \multicolumn{1}{l}{}   & \multicolumn{1}{l}{b} & Discover groups clustering of protected attribute in each ranking \\ 
    \midrule

    \multicolumn{1}{l}{G3} & \multicolumn{2}{l}{\textbf{Discover Bias in the Rankings}} \\ 
    \multicolumn{1}{l}{}   & \multicolumn{1}{l}{a} & {Discover (dis)advantaged groups in each ranking}    \\
    \multicolumn{1}{l}{}   & \multicolumn{1}{l}{b} & {Investigate the treatment of groups across rankings} \\
    \multicolumn{1}{l}{}   & \multicolumn{1}{l}{c} & {Intuit fairness of each ranking} \\
    \midrule

    \multicolumn{1}{l}{G4} & \multicolumn{2}{l}{\textbf{Generate Fair Consensus Rankings}} \\
    \multicolumn{1}{l}{}   & \multicolumn{1}{l}{a} & {Analyze multiple consensus rankings of different fairness level} \\
    \midrule

    \multicolumn{1}{l}{G5} & \multicolumn{2}{l}{\textbf{Discover Nuances (not captured by the model) }} \\
    \multicolumn{1}{l}{}   & \multicolumn{1}{l}{a} & {Analyze discrepancy on candidates between base rankings and consensus rankings} \\
    \multicolumn{1}{l}{}   & \multicolumn{1}{l}{b} & {Re-evaluate Fair Consensus Rankings} \\
    \bottomrule

  \end{tabular}
\end{table}

\begin{table}[tb]
  \caption{Activities resulting from the goals and activities analysis, designed to support the goals in Table \ref{table:goals-analysis}}
  \label{table:tasks-analysis}
  \centering
  \begin{tabular}{llp{.70\linewidth}}
    \toprule

    \multicolumn{3}{l}{\textbf{ACTIVITY}} \\ \midrule
    \multicolumn{1}{l}{G1a} & \multicolumn{2}{p{0.86\linewidth}}{\textbf{Discover (dis)agreement on each candidate between rankings}} \\
    \multicolumn{1}{l}{}   & \multicolumn{1}{l}{\label{task:A1}A1} & Locate each candidate across the rankings \\
    \multicolumn{1}{l}{}   & \multicolumn{1}{l}{\label{task:A2}A2} & Compare position of candidates across rankings \\
    \midrule

    \multicolumn{1}{l}{G1b} & \multicolumn{2}{p{0.86\linewidth}}{\textbf{Assess the discrepancy of candidates’ position between base rankings}}  \\
    \multicolumn{1}{l}{}   & \multicolumn{1}{l}{\label{task:A3}A3} & Compare position of multiple candidates between rankings \\
    \multicolumn{1}{l}{}   & \multicolumn{1}{l}{\label{task:A4}A4} & Compare Kendall Tau distance \cite{kendall1938new} between rankings \\
    \midrule

    \multicolumn{1}{l}{G2a} & \multicolumn{2}{p{0.86\linewidth}}{\textbf{Discover protected attribute groups of the candidates}}  \\
    \multicolumn{1}{l}{}   & \multicolumn{1}{l}{A5} & Identify protected attributes of candidates \\
    \midrule

    \multicolumn{1}{l}{G2b} & \multicolumn{2}{p{0.86\linewidth}}{\textbf{Discover groups clustering of protected attributes in each ranking}}  \\
    \multicolumn{1}{l}{}   & \multicolumn{1}{l}{A6} & Locate candidates of each group in a ranking     \\
    \multicolumn{1}{l}{}   & \multicolumn{1}{l}{A7} & Analyze distribution of candidates of each group     \\
    \midrule

    \multicolumn{1}{l}{G3a} & \multicolumn{2}{p{0.86\linewidth}}{\textbf{Discover (dis)advantaged groups in each ranking}}  \\
    \multicolumn{1}{l}{}   & \multicolumn{1}{l}{A8} & Identify FPR score of each group \\
    \multicolumn{1}{l}{}   & \multicolumn{1}{l}{A9} & Compare FPR score with a baseline fair score \\ \midrule

    \multicolumn{1}{l}{G3b} & \multicolumn{2}{p{0.86\linewidth}}{\textbf{Investigate the treatment of groups across rankings}}  \\
    \multicolumn{1}{l}{}   & \multicolumn{1}{l}{A10} & Compare FPR score of groups across rankings     \\ \midrule

    \multicolumn{1}{l}{G3c} & \multicolumn{2}{p{0.86\linewidth}}{\textbf{Intuit fairness of each ranking}}  \\
    \multicolumn{1}{l}{}   & \multicolumn{1}{l}{A11} & Identify ARP scores of the rankings    \\
    \multicolumn{1}{l}{}   & \multicolumn{1}{l}{A12} & Compare ARP across rankings \\ \midrule

    \multicolumn{1}{l}{G4a} & \multicolumn{2}{p{0.86\linewidth}}{\textbf{Analyze multiple consensus rankings of different fairness level}}  \\
    \multicolumn{1}{l}{}   & \multicolumn{1}{l}{A13} & Generate consensus rankings with different ARP thresholds     \\
    \multicolumn{1}{l}{}   & \multicolumn{1}{l}{A14} & Compare ARP and FPR scores between rankings (including consensus rankings)    \\
    \multicolumn{1}{l}{}   & \multicolumn{1}{l}{A15} & Compare Kendall Tau distance between rankings (including consensus rankings)   \\
    \midrule

    \multicolumn{1}{l}{G5a} & \multicolumn{2}{p{0.86\linewidth}}{\textbf{Analyze discrepancies on candidates between base rankings and
    consensus rankings}} \\
    \multicolumn{1}{l}{}   & \multicolumn{1}{l}{A16} & Compare individual candidate positions in base rankings with consensus rankings \\
    \multicolumn{1}{l}{}   & \multicolumn{1}{l}{A17} & Identify candidates with major differences in base rankings with consensus rankings     \\
    \midrule

    \multicolumn{1}{l}{G5b} & \multicolumn{2}{p{0.86\linewidth}}{\textbf{Re-evaluate Fair Consensus Rankings}}  \\
    \multicolumn{1}{l}{}   & \multicolumn{1}{l}{A18} & Manipulate candidate position or Re-iterate fair consensus ranking generation with different fairness threshold \\
    \bottomrule
  \end{tabular}
\end{table}

\subsection{\fairfuse}

The \fairfuse system \hs{(\autoref{fig:teaser}B)} consists of several views to support the goals (Table \ref{table:goals-analysis}) and activities (Table \ref{table:tasks-analysis}).

\begin{itemize}
    \item \textbf{Ranking Exploration View} uses parallel coordinates plot to explore and compare rankings of candidates between multiple stakeholders (A1, A2, A3, A16, A17) as shown in Figure \ref{fig:fairfuse-system}F. Each candidate's set of attributes and values are represented by glyphs and colors \cite{maguire2012taxonomy} (A5, A6), collectively called  a Candidate Card. By dragging-and-dropping the Candidate Card (A18), any generated consensus ranking can be adjusted if necessary.
    \item \textbf{Group Fairness View} (Figure \ref{fig:fairfuse-system}E) captures fairness of a ranking at individual group level utilizing FPR score \cite{cachel2022manirank} (A6, A7, A8, A9, A10) and holistically across groups in the ranking using ARP score \cite{cachel2022manirank} (A11, A12, A14).
    \item \textbf{Similarity View} (Figure \ref{fig:fairfuse-system}B) uses a heatmap to show the similarity between any two rankings (A4, A15) with darker squares representing higher similarity between the rankings. This includes the ability to compare similarities between any two base rankings, and a base ranking with a consensus ranking. The similarity measure is calculated using a common measure for rank dissimilarity called Kendall-Tau distance \cite{kendall1938new}.
    \item \textbf{Ranking Generation} process uses a button to 
    first generate a consensus ranking without any fairness intervention
    \ref{fig:fairfuse-system}A. After the consensus ranking is displayed, the generation button is replaced with a slider (Figure \ref{fig:fairfuse-system}E) -- allowing the fairness threshold of generated consensus ranking to be adjusted (A13, A18). This process utilizes the Fair-Copeland algorithm \cite{cachel2022manirank}.
\end{itemize}

\subsection{\confuse}

An alternate \hs{and functionally equally capable }version of \fairfuse was created, called \confuse \hs{(\autoref{fig:teaser}A)}, which acts as a baseline for comparison in our study.
Changes included the removal of 1) encodings of fairness metrics in the Group Fairness View (Figure \ref{fig:fairfuse-system}H), 2) the Similarity View which uses metrics to compare the similarity of fair rankings, and 3) the fairness algorithm in the consensus ranking generation process, which had a slider to control the ARP \cite{cachel2022manirank}. Differences are shown in Figure \ref{fig:fairfuse-system}.

\section{Study Design}

\begin{figure*}[tb]
    \includegraphics[width=\textwidth]{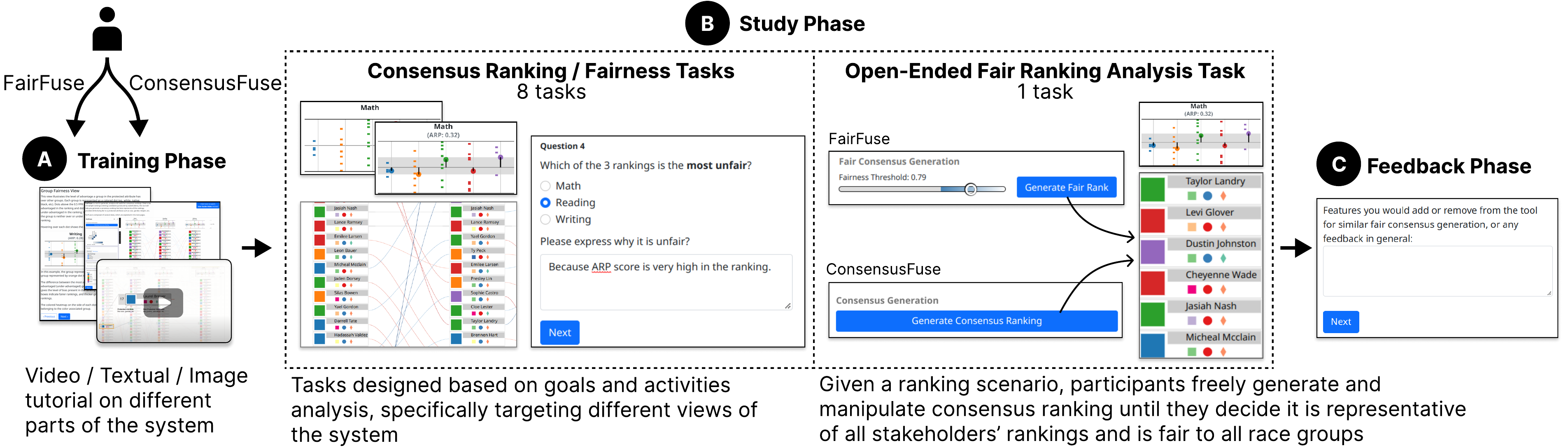}
    \caption{Study Design: We explore  using visualization-enabled fairness metrics in building a fair consensus ranking. Participants are divided into two conditions, \fairfuse: system with visualization-enabled fairness metrics and \confuse: system without visualizations for fairness metrics. Participants go through three phases. A) Training Phase, B) Study Phase and C) General Feedback Phase.
    }
    \Description{Study Design}
    \label{fig:study-design}
\end{figure*}

We aim to investigate the challenges and opportunities of a system like \fairfuse for the activities associated with fairness-oriented tasks. In our study, we presented a scenario where participants were tasked with constructing a fair consensus ranking for scholarship distribution based on teachers' rankings of students.
We performed a between-subjects study in which each participant was assigned to use either \fairfuse or \confuse system.

\subsection{Procedure}
We recruited 200 English-speaking participants agreeing to an IRB-approved consent form on Prolific, a crowd-sourcing platform. Based on multiple pilot studies, each participant was paid \$5 USD for an estimated 25-minute study time, with an hourly rate of \$12.00 USD.
Since both the system used for the study is built for large screens, participants were filtered to use only desktop devices using Prolific's screening process.
Our study consists of 3 phases: training, study, and feedback phase (Figure \ref{fig:study-design}) as seen in a similar user study in the literature \cite{nobre2020evaluating}.

{\bf Training Phase.} The study starts with the training phase (Figure \ref{fig:study-design}A) introducing participants to different parts of the system through textual, figurative, and video explanations while also 
encourging
them with analysis regarding consensus finding and bias mitigation.

{\bf Study Phase.} The second phase (Figure \ref{fig:study-design}B) involved participants completing tasks. Both \fairfuse and \confuse systems' interfaces were adjusted to include a view displaying the tasks. The sidebar was shortened to accommodate the tasks and participant answers at the bottom.
During this phase, the participants interacted with the visualizations to find the answer(s). Each task was followed by a multiple-choice form with a dropdown or checkbox, and some were also followed by a free text form.
The tasks in this phase were designed to increase in complexity gradually. Participants could refer back to the tutorial if they encountered difficulty.
This phase was further divided into two parts. The first part focused on the systems' specific views and activities (Table \ref{table:tasks-analysis}) while additionally serving as a guided tutorial for the second part of this phase. On the other hand, the second part invited participants to interact with all system views while completing an open-ended task of constructing a fair consensus ranking.

{\bf Feedback Phase.} The final phase of the study (Figure \ref{fig:study-design}C) collected qualitative feedback on the system regarding generating a fair consensus ranking and demographics-related information.

\subsection{Tasks Scenario Data}

For this study, we adapted the data from the publicly available dataset \cite{kimmons} of students' rankings. 
The dataset contains multiple attributes, but for generating a consensus ranking, we used the relative ordering of students in three subjects, math, reading, and writing, as base rankings. 
Since our study phase has two parts, we 
created two datasets of 30 students each, where one dataset was used for each of the two study parts. 
The dataset was split such that both had all 5 groups of the protected attribute, race, the advantaged and disadvantaged groups can be separable. 
Race was the protected attribute for both datasets, with five groups: White, Native, Black, Asian, and Pacific Islander. 

\subsection{Study Task Design}

Targeting the goals and activities (Table \ref{table:tasks-analysis}), we created a set of tasks for the participants, listed in Table \ref{tab:questions-participants}.
The first eight tasks focus on different individual views of the system. These tasks encompass the Ranking Exploration View with candidate cards containing attributes of the candidate and parallel coordinates plot of the rankings, Similarity View, Group Fairness View, and the Consensus Generation process. The final task asks participants to conduct a free-form fair consensus ranking generation.

\begin{table*}[!t]
  \caption{List of task prompts given to the participants. Tasks are targeted at the \textit{Goals and Activities} (Table \ref{table:tasks-analysis}) analysis. }

  \label{tab:questions-participants}
  \centering
  \begin{tabular}{cp{.28\textwidth}p{.4\textwidth}l}
  \toprule
  & \bfseries Task & \bfseries Task Prompt & \bfseries Target Activity\\
  \midrule

    \bfseries T1 & Locating protected attribute  & What is the race of Taylor Landry? & A1, A5 \\
    \bfseries T2 & Identifying Advantaged Group(s) & Which race groups are over advantaged in the Math ranking?
    & A6, A7, A8, A9\\
    \bfseries T3 & Visualization Use & {Click on the visualization you primarily used to deduce the answer for the previous question?
    } & \\
    
    \bfseries T4 & Identifying Attribute-level Unfairness & Which of the 3 rankings is the most unfair? Please express why it is unfair? & A11, A12\\
    \bfseries T5 & Identifying Group-level Unfairness & Which of the 3 rankings have the most advantaged White race group? & A6, A7, A10\\
    \bfseries T6 & Utilizing PCP Position Comparison & How is Taylor Landry's position ranked in Math compared to Reading? &  A2, A3\\
    \bfseries T7 & Interpreting PCP Gradient & Select the candidate with most disagreement between Math and Reading rankings. 
    Please explain how you deduced your answer. & A2\\
    \bfseries T8 & Using Consensus Generation Procedure & STEP 1: Generate a consensus ranking using the button on the top of the left sidebar.

STEP 2: Use the pin icon in the heading of the generated ranking to pin the ranking.

STEP 3: Please identify which base ranking is most dissimilar to consensus ranking you just generated. & A4, A13, A15\\
    
    \hline
    \bfseries T9 & Using Fair Consensus Generation Procedure & Generate a fair consensus ranking that:

1. Is representative of all the base rankings\newline
2. Does not over or under advantage race groups & A1 - A18 \\

  \bottomrule
  \end{tabular}
\end{table*}
\section{Results}


We recruited 200 participants (a number obtained via power analyses following pilot studies) and evenly divided them into two groups, namely,  \fairfuse and \confuse. 
\hs{Random assignment was achieved through round-robin online recruitment using the Prolific platform. 
Prolific reporting shows that 99 participants (separate from the 200 completions) returned the experiment before completing it.
(On Prolific, participants can discontinue the experiment for any reason.)
Beyond these, 6 participants in total timed out.}
We computed $95\%$ confidence interval using a bootstrapped method and effect size using Cohen's \textit{d}. Our results also include p-value (p) from the Wilcox Test (W).

\subsection{Ranking Exploration Tasks}
\begin{figure*}[tb]
    \begin{subfigure}[t]{0.32\linewidth}
        \includegraphics[width=\linewidth]{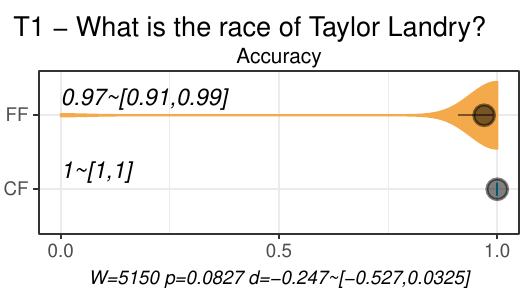}
        \caption{Locating protected attribute [T1]}
        \label{fig:violin-ci-1}
    \end{subfigure}
    \hfill
    \begin{subfigure}[t]{0.32\linewidth}
        \includegraphics[width=\linewidth]{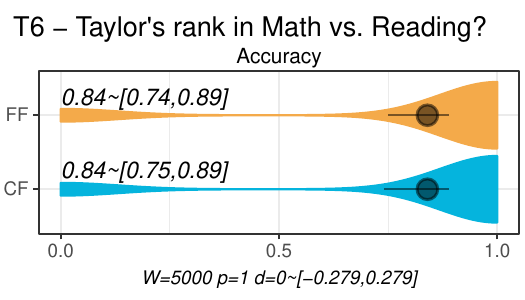}
        \caption{Utilizing PCP Position Comparison [T6]}
        \label{fig:violin-ci-6}
    \end{subfigure}
    \hfill
    \begin{subfigure}[t]{0.32\linewidth}
        \includegraphics[width=\linewidth]{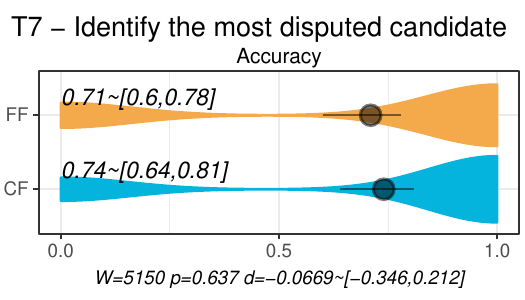}
        \caption{Interpreting PCP Gradient [T7]}
        \label{fig:violin-ci-7}
    \end{subfigure}
  \caption{Results for Ranking Exploration tasks}
  \label{fig:violin-ci-ranking-exploration}
\end{figure*}
Since three of the tasks, \textbf{T1}, \textbf{T6} and \textbf{T7}, relied on the unmodified views presented for both groups, we observe that there is no significant difference
in the answers given by the participants. The violin plot with confidence intervals, p-value and effect size are shown in Figure \ref{fig:violin-ci-ranking-exploration}. We report 
no significant difference in all three tasks between the two conditions, namely, $p=0.0827$, $p=1.0$, and $p=0.637$, respectively.
We find that participants are able to identify attributes and compare positions of candidates between rankings using Parallel Coordinates Plot in both systems. 
For \textbf{T7}, which is an advanced task compared to \textbf{T1} and \textbf{T6}, we see a slight decrease in the correct answers. \textbf{T7} asked participants to identify the candidate with the most disagreement between two rankings. This task involved identifying a candidate card connected with a line between two adjacent rankings with the most inclination. 

\subsection{Fairness-oriented Tasks}
\begin{figure}[h]
    \begin{subfigure}[t]{\linewidth}
        \includegraphics[width=\linewidth]{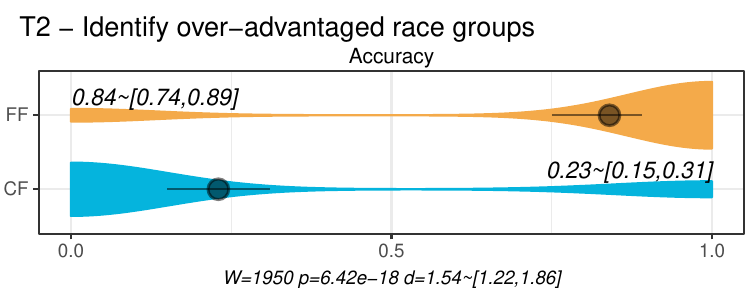}
        \caption{Results for Identifying advantaged Group(s) [T2] (both native and white as correct answer)}
        \label{fig:violin-ci-2}
    \end{subfigure}
    \par\bigskip
    \begin{subfigure}[t]{\linewidth}
        \includegraphics[width=\linewidth]{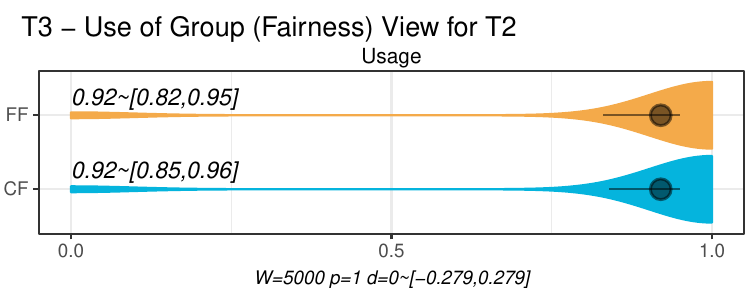}
        \caption{Results for use of Group (Fairness) View for T2 [T3]}
        \label{fig:violin-ci-3}
    \end{subfigure}
    \par\bigskip
    \begin{subfigure}[t]{0.48\textwidth}
        \includegraphics[width=\linewidth]{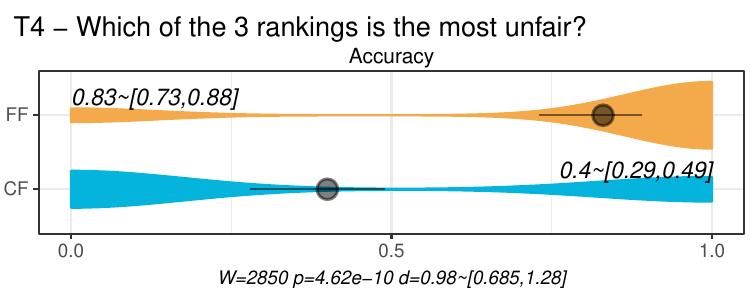}
        \caption{Results for Identifying Attribute-level Unfairness [T4]}
        \label{fig:violin-ci-4}
    \end{subfigure}
    \par\bigskip
    \begin{subfigure}[t]{0.48\textwidth}
        \includegraphics[width=\linewidth]{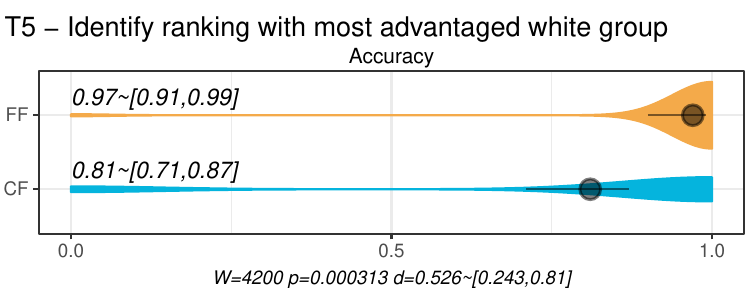}
        \caption{Results for Identifying Group-level Unfairness [T5]}
        \label{fig:violin-ci-5}
    \end{subfigure}
    \caption{Results for Fairness-oriented tasks}
    \label{fig:violin-ci-fairness}
\end{figure}

\textbf{T2} asks participants to identify advantaged groups in one of the three rankings provided. During the experiment, participants were provided with checkboxes of five race groups allowing them to select multiple race groups. The ground truth included two advantaged race groups based on the FPR scores. We observe that the user performance in \fairfuse ($M = 0.84\sim [0.74, 0.89]$) is significantly better than \confuse ($M = 0.23\sim[0.15, 0.31$]) as shown in the violin plot (Figure \ref{fig:violin-ci-2}) with a large effect size ($d=1.54\sim[1.22,1.86]$). The careful design of the Group Fairness View in \fairfuse with the affordance of a horizontal line providing a visual cue of the baseline that separates the advantaged from disadvantaged groups could have helped \fairfuse achieve  better accuracy for this question. We also find that both \fairfuse and \confuse participants use the same view for tackling this question \textbf{T2} as seen in Figure \ref{fig:violin-ci-3}.

\begin{figure}[h]
    \begin{subfigure}[t]{\linewidth}
        \includegraphics[width=\textwidth]{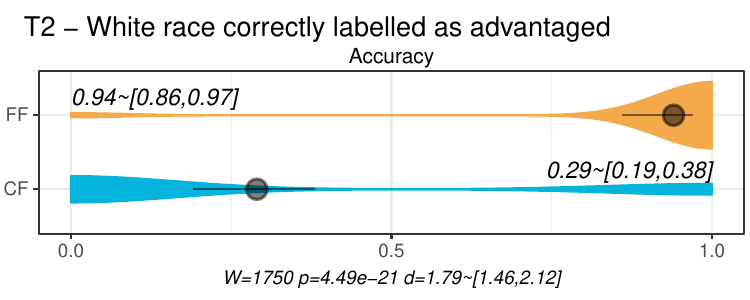}
        \caption{Results for Identifying Advantaged Group(s) with \textbf{white} as correct answer [T2]}
        \label{fig:violin-ci-2-white}
    \end{subfigure}
    \par\bigskip
    \begin{subfigure}[t]{\linewidth}
        \includegraphics[width=\textwidth]{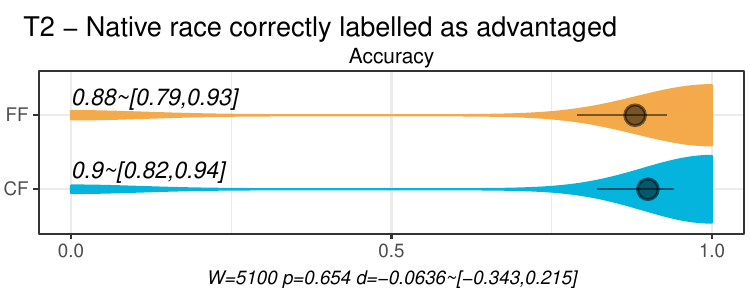}
        \caption{Results for Identifying Advantaged Group(s) with \textbf{native} as correct answer [T2]}
        \label{fig:violin-ci-2-native}
    \end{subfigure}
    \caption{Results per individual group for Identifying advantaged Group(s) [T2]}
    \label{fig:violin-ci-2-all}
\end{figure}

It's noteworthy that the majority of participants in the \confuse study selected one of the two correct advantaged groups, while the participants in \fairfuse identified both correct advantaged groups (as shown in Figures \ref{fig:violin-ci-2-white} and \ref{fig:violin-ci-2-native}). This highlights the significance of fairness metrics and visualizations in identifying multiple advantaged or disadvantaged groups when a large number of groups are involved.

For \textbf{T4}, participants were asked to identify the most unfair ranking among the three rankings provided. While \textbf{T2} focused on the level of advantage each group has using FPR measure \cite{cachel2022manirank}, \textbf{T4} focused on utilizing the ARP metric \cite{cachel2022manirank}. Similar to \textbf{T2}, with \textbf{T4}, we get significantly different results between the two conditions with  high accuracy in \fairfuse (\confuse: M = 0.4 [0.29, 0.49] vs. \fairfuse: M = 0.83 [0.73, 0.88]) as shown in Figure \ref{fig:violin-ci-4}. We also instructed participants to express why they think their ranking choice is unfair. 
\hs{Participant comments reflect at least two types of reasoning in the \fairfuse condition}: expression at the vis-level and expression at the understanding level.
Expression at vis-level reports the ARP score or visualization that mimics the ARP score, such as:
\begin{displayquote}
    \textit{The grey bar is the widest with [the] highest ARP index.}
\end{displayquote}
Expression at the understanding level goes beyond just reporting the ARP score, such as: 
\begin{displayquote}
    \textit{Reading shows the largest disparity between the highest and lowest group fairness scores, ergo the disparity between highs and lows would be the most unfair.}
\end{displayquote}

In contrast, some \confuse participants considered only a single group resulting in incorrect answers, such as:
\begin{displayquote}
    \textit{The black group is very under-advantaged and is ranked a lot lower than other groups.}
\end{displayquote}

Also, it is interesting that some \confuse participants did meticulous calculations of individual groups, such as: 

\begin{displayquote}
    \textit{100\% of the white students are in the top half, but only 28.5\% of the black students are.}
\end{displayquote}

Task \textbf{T5} builds from \textbf{T2} and \textbf{T4}, where participants were asked to identify the ranking with the most advantaged White race group. 
We find a small but significant difference in accuracy ($p = 0.000313$; \fairfuse: $M = 0.97\sim[0.91, 0.99]$ vs. \confuse: $M = 0.81\sim[0.71, 0.87]$) with medium effect size ($d = 0.526\sim[0.243,0.81]$) as shown in Figure \ref{fig:violin-ci-5}.
This may be because \textbf{T5} specifically asks about a particular group instead of multiple groups resulting in similar results like Identifying Advantaged Group(s) (\textbf{T2}) with native as a correct answer (Figure \ref{fig:violin-ci-2-native}).

\subsection{Consensus Representation and Analysis Tasks} \label{sec:consensus-representation}

\begin{figure}[h]
    \centering
    \includegraphics[width=\linewidth]{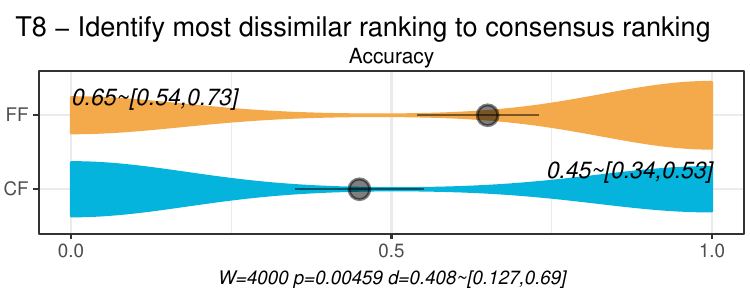}
    \caption{Results on Using Consensus Generation Procedure [T8]}
    \label{fig:violin-ci-8}
\end{figure}

To assess \fairfuse's performance in identifying similarity of consensus ranking to base rankings, we device \textbf{T8}.
To ensure a fair comparison between the systems, we asked both groups to start with generating a fairness-unaware consensus ranking, followed by selecting the most dissimilar base ranking. 
This way, both groups have the same state of rankings to begin with.
The violin plot shows the result (Figure \ref{fig:violin-ci-8}) with a significant difference between the two groups and a medium effect size ($p=0.00459$, $d=0.408\sim[0.127,0.69]$). Although \fairfuse ($M = 0.65\sim[0.54,0.73]$) was more accurate than \confuse ($M = 0.45\sim[0.34,0.53]$), the advantage was not very high. 

\hs{For this particular task, we also asked participants elaborate on their answers. We find that some participants in \fairfuse, even though they correctly identify the most dissimilar ranking, mention using the Group Fairness View instead of the Similarity View, such as:}

\begin{displayquote}
   \textit{The ARP of reading is the furthest away from the ARP of the consensus.}
\end{displayquote}

Interestingly, despite having the Similarity View in \fairfuse, some participants either used a process similar to that of the \confuse participants by dragging individual base rankings towards the consensus ranking and counting line crossings, or didn't find the view useful.

\begin{displayquote}
    \textit{I dragged each individual ranking over to place it side-by-side with the consensus ranking.
    [...]
    reading had the most lines that strayed from this path.}
\end{displayquote}
\begin{displayquote}
    \textit{I did not find the Similarity View very helpful.}
\end{displayquote}

\hs{As a result, while quantitative data in aggregate supports the notion that \fairfuse performs better in identifying the (dis)similarity between the consensus ranking and base rankings, qualitative results do not fully support this conclusion. Participants may focus more on the fairness metrics compared to other available information like the Similarity View.}

\subsection{Open-Ended Fair Ranking Analysis Task}

\begin{figure*}[h]
    \begin{subfigure}[t]{0.32\linewidth}
        \includegraphics[width=\linewidth]{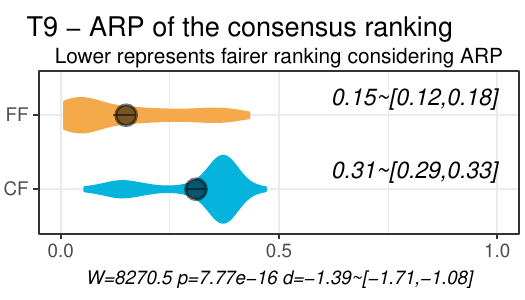}
        \caption{Results for ARP of generated fair consensus ranking [T9]}
        \label{fig:violin-ci-10-arp}
    \end{subfigure}
    \hfill
    \begin{subfigure}[t]{0.32\linewidth}
        \includegraphics[width=\linewidth]{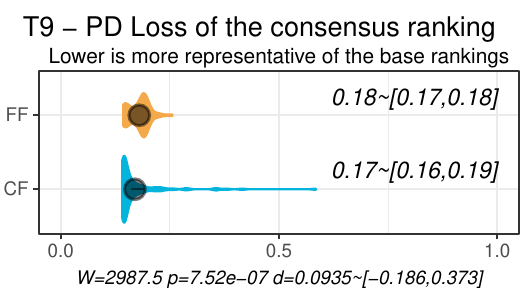}
        \caption{Results for PD Loss of generated fair consensus ranking [T9]}
        \label{fig:violin-ci-10-pdloss}
    \end{subfigure}
    \hfill
    \begin{subfigure}[t]{0.32\linewidth}
        \includegraphics[width=\linewidth]{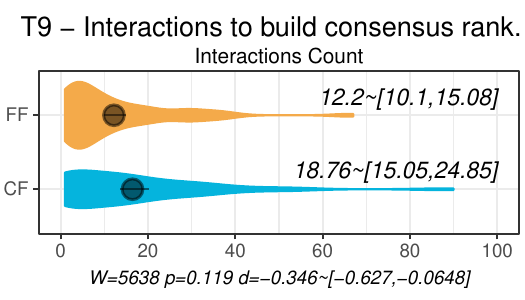}
        \caption{Total interactions the participants made to build fair consensus ranking [T9]
        }
        \label{fig:violin-ci-10-interactions}        
    \end{subfigure}
    \label{fig:violin-ci-h4}
    \caption{Results for Open-Ended Fair Ranking Analysis Task}
\end{figure*}

In \textbf{T9}, we ask participants to generate a fair consensus ranking that is representative of all the base rankings such that it does not over or under-advantage race groups. 
We analyze the ARP scores between the two groups (which ranges from $0$ to $1$, with $0$ representing a ranking with perfect statistical parity \cite{cachel2022manirank}) to measure the group fairness requirement.
We find that \fairfuse participants generally agree on consensus rankings with lower ARP scores ($M=0.15\sim[0.12,0.18]$) compared to \confuse ($M=0.31\sim[0.29,0.33]$) with a large effect size ($d=-1.39\sim[-1.71,-1.08]$), interpreting that the participants fail to create a fairer consensus ranking in \confuse. However, we note that some of the participants, even without the fairness metrics and its visualizations, built consensus rankings with low ARP scores.

\begin{figure}[tb]
  \centering
  \includegraphics[width=\linewidth]{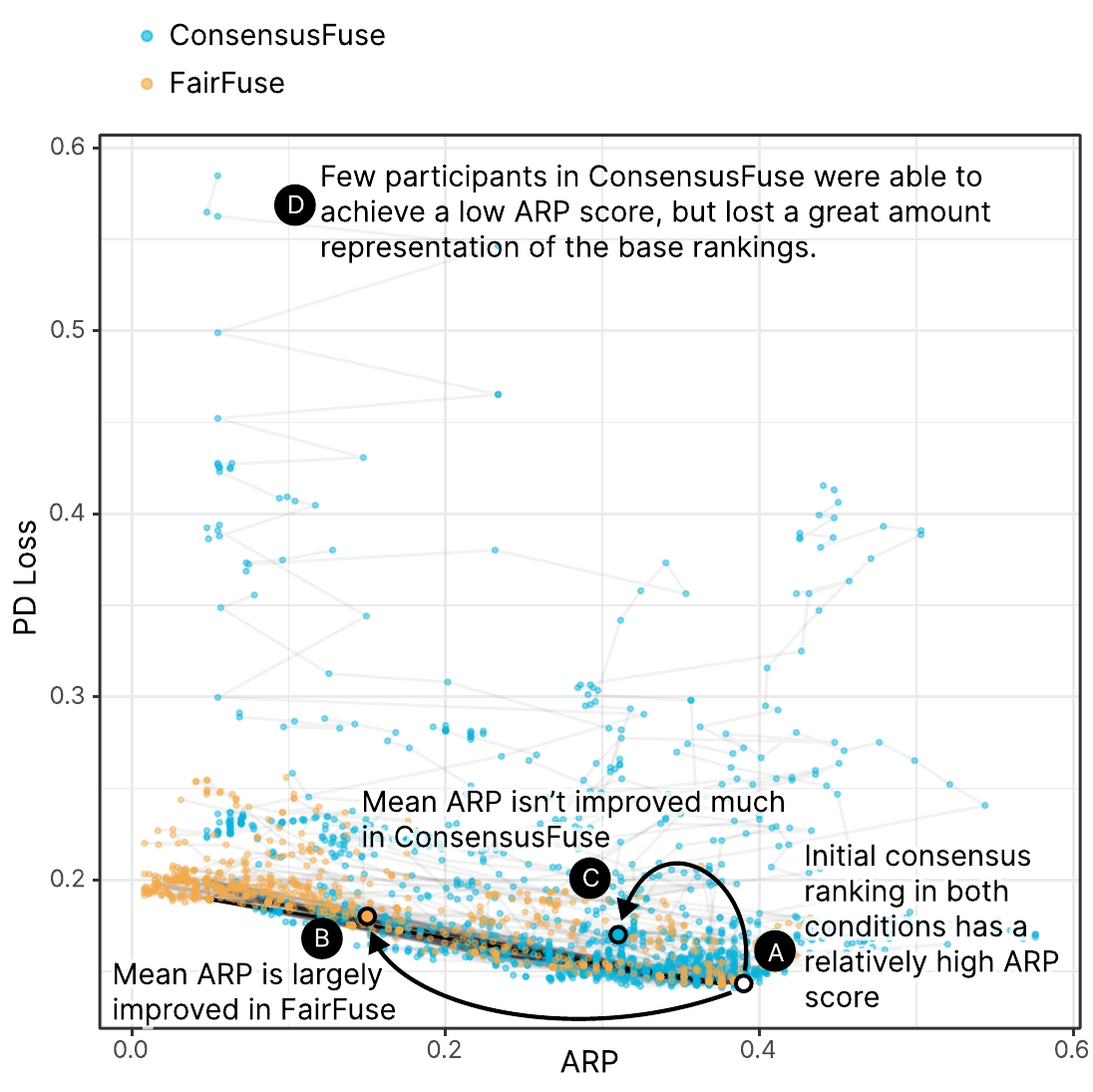}
  \caption{Results of ARP vs. PD Loss throughout each user interaction while generating a fair consensus ranking.
  The white dot indicates the ARP and PD Loss of the initial consensus ranking in both conditions.}
  \label{fig:scatter-arp-pdloss}
\end{figure}

We observe that the PD Loss \cite{cachel2022manirank} (representation of base rankings in the consensus ranking, with $0$ representing that all the base rankings exactly match the consensus ranking) in both groups are similar (Figure \ref{fig:violin-ci-10-pdloss}) despite some participants in \confuse ending up producing rankings that are far in distance from the base rankings, yet on the fair side, as seen in scatterplot (Figure \ref{fig:scatter-arp-pdloss}D).
Figure \ref{fig:scatter-arp-pdloss}A marks the initial consensus ranking for both conditions, which has a relatively higher ARP score. 
Interactions included drag-and-drops of candidate cards for updating the consensus ranking and generation of consensus rankings.
Figure \ref{fig:scatter-arp-pdloss}B marks the vastly improved mean ARP value in \fairfuse compared to Figure \ref{fig:scatter-arp-pdloss}C in \confuse. 
We find that \fairfuse participants make fewer interactions to agree on a fair consensus ranking as shown in the violin plot (Figure \ref{fig:violin-ci-10-interactions}) (\confuse: $M = 18.76\sim[15.03, 24.62]$ vs. \fairfuse: $M = 12.2\sim[10.1, 15.08]$).

\section{Discussion}

Overall results suggest that while both systems are suited for exploring ranking-related tasks, \fairfuse outperforms in terms of accuracy in fairness-related tasks.
Also,  fewer interactions are involved in generating 
fair consensus rankings in   \fairfuse.
We find that \fairfuse, with its unique visualization-enabled fairness metrics, helps keep a balance between generating a ranking that maximizes the agreement of base rankings while keeping it as fair as possible concerning statistical parity, a common definition of fairness.
However, we also find that users are drawn towards relying on fairness metrics and algorithms to complete the tasks, sometimes erroneously so. 
This introduces a tension between the goals of building a representative consensus versus ensuring that it is fair-- a tension that creates interesting constraints and challenges for design. 

Based on our study, we distilled a set of \textbf{4} takeaways summarizing how we observed visualized fairness metrics and algorithms helping or hindering in tasks and decision-making contexts. 
These takeaways may hold broader implications for developers of fairness metrics and algorithms, designers of visual interfaces, and the fairness community at large.

\subsection{Help: Researchers developing fairness-aware algorithms should incorporate ways for end-users to tune fairness, relative to other problem objectives}

Given evolving societal norms and values, definitions of fairness can change over time and place. Definitions also vary from one discipline to another \cite{mulligan2019thing}.
Algorithms designed to assist in incorporating fairness incorporate ways for decision-makers to tune fairness in the specific problem context. This increases both agency on the part of decision-makers, and incorporates their specific domain knowledge and worldviews. 
While \fairfuse could produce an absolute fair consensus ranking based on the algorithm used, we find that participants set the fairness threshold close to the absolute threshold to generate a fair consensus ranking. 
This behavior suggests that allowing individuals to adjust the parameters of an algorithm can lead to more satisfactory and appropriate results. 
Moreover, making fair algorithms tunable allows for more transparency and accountability in decision-making, as decision-makers can see and understand the factors influencing the algorithm's output.

\subsection{Hinder: Visualization designers should be mindful that visually displaying fairness metric may lead to increased credence in
and over-reliance on metrics}

Our results suggest that Decision makers tend to make decisions that are consistent with visualization-enabled fairness metrics \autoref{fig:scatter-arp-pdloss}.
From a positive perspective, alignment with fairness metrics can promote fairness in decision-making. Yet, designers should also be cautious about the consequences of such drift. Nudging decision-makers toward visual indications of fairness may result in decision-makers blindly trusting such metrics and algorithms could miss the societal nuances that the metrics cannot capture, which is reflected in participants' comments, \eg: 
\textit{"Fairness threshold [of] 1 seems to do the job?"}, \textit{"I use the slider and slide it to fairness threshold to 1. 
   [...]
   Then the ranking will be unbiased."}
Visualization designers and the fairness community should be mindful of the potential for ``fairness drift'', particularly as metrics are increasingly incorporated into visual interfaces.

\subsection{Help: Properly designed visualizations of fairness metrics can help people navigate complexity in decision-making contexts}

The multi-objective nature of fairness related tasks can be tricky to navigate for non-expert users where achieving a goal (such as a building a good consensus ranking) is also subjected to bias mitigation. Inclusion of large number variables can make it worse as we see in our results where participants were able to identify only one of the two advantaged groups without the help of visualizations supporting fairness metrics (Figure \ref{fig:violin-ci-2-all}). Identifying such groups can highlight areas of concern, making it easy for further analysis in mitigating bias. Properly designed visualization of fairness metrics can help identify bias across a larger number of variables helping individuals to make informed decisions in the decision-making process. 

\subsection{Hinder: Improperly designed fairness metrics visualizations can lead people to incorrect conclusions}

While visualization tools like \fairfuse can be used to promote fairness in building a consensus ranking,  it is crucial for visualization designers to be mindful of the way in which fairness metrics are presented, as improper design can lead individuals to draw incorrect conclusions. 
For example, in the case of \fairfuse, 
presenting new visualization such as the Group Fairness View on occasion led participants to overlook other important information such as the Similarity View (see Section \ref{sec:consensus-representation}),
Yet, the later is equally important in maintaining consensus.
Failure to do so could result in an incomplete understanding of the task at hand. 

\section{Limitations and Future Work}

The limitations of \fairfuse \cite{shrestha2022fairfuse} are also relevant to this crowd-sourced study. \fairfuse focusses on ARP and FPR fairness metrics \cite{cachel2022manirank} within the widely accepted definition of group fairness in the fairness community. It also considers one tunable algorithm for generating fair consensus ranking. The fulfillment of the goals of the system relied on those metrics in the tasks abstraction phase. 
However, Verma and Rubin \cite{verma2018fairness} highlight that a decision considered fair by one definition may be deemed unfair by others, and laypeople's judgment often aligns with simple notions of fairness like group fairness \cite{constantin2022algorithmic}.
Therefore, future work could incorporate multiple fairness definitions and conduct similar user studies. 
Future studies might also examine the potential benefits and drawbacks of using tunable algorithms like in \fairfuse for fairness-related tasks. In addition, these studies could assess the impact on decision-makers trust in these systems and the possibility of an increased cognitive load.

\section{Conclusion}
The concern for fairness in AI tools and online platforms has amplified the need for effective methods of identifying and mitigating bias in ranking processes. However, the complexities of fair consensus ranking, including multiple bias-causing factors and nuanced ethical and societal values, make a fully automated system unreliable. Human-in-the-loop systems, which offer a comprehensive approach to bias mitigation, can be valuable, but there is limited evidence on the benefits and challenges of designing visualizations that support fairness metrics.

To investigate these challenges, we conducted a crowd-sourced study across goals and activities designed for building a fair consensus ranking between a metrics-based visualization \fairfuse and a non-metric based visualization system \confuse.
Our findings suggest that well-designed visualizations can aid in creating fair consensus rankings, but they may also  hinder certain tasks, particularly balancing goals beyond fairness in decision-making contexts.


\begin{acks}
This work was supported in part by NSF IIS \#2007932.
\end{acks}

\bibliographystyle{ACM-Reference-Format}
\bibliography{manuscript}


\begin{thebibliography}{58}


\ifx \showCODEN    \undefined \def \showCODEN     #1{\unskip}     \fi
\ifx \showDOI      \undefined \def \showDOI       #1{#1}\fi
\ifx \showISBNx    \undefined \def \showISBNx     #1{\unskip}     \fi
\ifx \showISBNxiii \undefined \def \showISBNxiii  #1{\unskip}     \fi
\ifx \showISSN     \undefined \def \showISSN      #1{\unskip}     \fi
\ifx \showLCCN     \undefined \def \showLCCN      #1{\unskip}     \fi
\ifx \shownote     \undefined \def \shownote      #1{#1}          \fi
\ifx \showarticletitle \undefined \def \showarticletitle #1{#1}   \fi
\ifx \showURL      \undefined \def \showURL       {\relax}        \fi
\providecommand\bibfield[2]{#2}
\providecommand\bibinfo[2]{#2}
\providecommand\natexlab[1]{#1}
\providecommand\showeprint[2][]{arXiv:#2}

\bibitem[Ahn and Lin(2019)]%
        {ahn2019fairsight}
\bibfield{author}{\bibinfo{person}{Yongsu Ahn} {and} \bibinfo{person}{Yu-Ru
  Lin}.} \bibinfo{year}{2019}\natexlab{}.
\newblock \showarticletitle{Fairsight: Visual analytics for fairness in
  decision making}.
\newblock \bibinfo{journal}{\emph{IEEE transactions on visualization and
  computer graphics}} \bibinfo{volume}{26}, \bibinfo{number}{1}
  (\bibinfo{year}{2019}), \bibinfo{pages}{1086--1095}.
\newblock


\bibitem[Bajracharya et~al\mbox{.}(2018)]%
        {bajracharya2018interactive}
\bibfield{author}{\bibinfo{person}{S Bajracharya}, \bibinfo{person}{Giuseppe
  Carenini}, \bibinfo{person}{B Chamberlain}, \bibinfo{person}{K Chen},
  \bibinfo{person}{D Klein}, \bibinfo{person}{David Poole},
  \bibinfo{person}{Hamed Taheri}, {and} \bibinfo{person}{Gunilla {\"O}berg}.}
  \bibinfo{year}{2018}\natexlab{}.
\newblock \showarticletitle{Interactive visualization for group decision
  analysis}.
\newblock \bibinfo{journal}{\emph{International Journal of Information
  Technology \& Decision Making}} \bibinfo{volume}{17}, \bibinfo{number}{06}
  (\bibinfo{year}{2018}), \bibinfo{pages}{1839--1864}.
\newblock


\bibitem[Bantilan(2018)]%
        {bantilan2018themis}
\bibfield{author}{\bibinfo{person}{Niels Bantilan}.}
  \bibinfo{year}{2018}\natexlab{}.
\newblock \showarticletitle{Themis-ml: A fairness-aware machine learning
  interface for end-to-end discrimination discovery and mitigation}.
\newblock \bibinfo{journal}{\emph{Journal of Technology in Human Services}}
  \bibinfo{volume}{36}, \bibinfo{number}{1} (\bibinfo{year}{2018}),
  \bibinfo{pages}{15--30}.
\newblock


\bibitem[Behrisch et~al\mbox{.}(2013)]%
        {behrisch2013visual}
\bibfield{author}{\bibinfo{person}{Michael Behrisch}, \bibinfo{person}{James
  Davey}, \bibinfo{person}{Svenja Simon}, \bibinfo{person}{Tobias Schreck},
  \bibinfo{person}{Daniel Keim}, {and} \bibinfo{person}{J{\"o}rn Kohlhammer}.}
  \bibinfo{year}{2013}\natexlab{}.
\newblock \showarticletitle{Visual comparison of orderings and rankings}. In
  \bibinfo{booktitle}{\emph{EuroVis}}.
\newblock


\bibitem[Bellamy et~al\mbox{.}(2018)]%
        {bellamy2018ai}
\bibfield{author}{\bibinfo{person}{Rachel~KE Bellamy}, \bibinfo{person}{Kuntal
  Dey}, \bibinfo{person}{Michael Hind}, \bibinfo{person}{Samuel~C Hoffman},
  \bibinfo{person}{Stephanie Houde}, \bibinfo{person}{Kalapriya Kannan},
  \bibinfo{person}{Pranay Lohia}, \bibinfo{person}{Jacquelyn Martino},
  \bibinfo{person}{Sameep Mehta}, \bibinfo{person}{Aleksandra Mojsilovic},
  {et~al\mbox{.}}} \bibinfo{year}{2018}\natexlab{}.
\newblock \showarticletitle{AI Fairness 360: An extensible toolkit for
  detecting, understanding, and mitigating unwanted algorithmic bias}.
\newblock \bibinfo{journal}{\emph{arXiv preprint arXiv:1810.01943}}
  (\bibinfo{year}{2018}).
\newblock


\bibitem[Bird et~al\mbox{.}(2020)]%
        {bird2020fairlearn}
\bibfield{author}{\bibinfo{person}{Sarah Bird}, \bibinfo{person}{Miro
  Dud{\'\i}k}, \bibinfo{person}{Richard Edgar}, \bibinfo{person}{Brandon Horn},
  \bibinfo{person}{Roman Lutz}, \bibinfo{person}{Vanessa Milan},
  \bibinfo{person}{Mehrnoosh Sameki}, \bibinfo{person}{Hanna Wallach}, {and}
  \bibinfo{person}{Kathleen Walker}.} \bibinfo{year}{2020}\natexlab{}.
\newblock \showarticletitle{Fairlearn: A toolkit for assessing and improving
  fairness in AI}.
\newblock \bibinfo{journal}{\emph{Microsoft, Tech. Rep. MSR-TR-2020-32}}
  (\bibinfo{year}{2020}).
\newblock


\bibitem[Borda et~al\mbox{.}(1781)]%
        {borda1781mathematical}
\bibfield{author}{\bibinfo{person}{Jean-Charles~de Borda} {et~al\mbox{.}}}
  \bibinfo{year}{1781}\natexlab{}.
\newblock \showarticletitle{Mathematical derivation of an election system}.
\newblock \bibinfo{journal}{\emph{Isis}} \bibinfo{volume}{44},
  \bibinfo{number}{1-2} (\bibinfo{year}{1781}), \bibinfo{pages}{42--51}.
\newblock


\bibitem[Brehmer and Munzner(2013)]%
        {brehmer2013multi}
\bibfield{author}{\bibinfo{person}{Matthew Brehmer} {and}
  \bibinfo{person}{Tamara Munzner}.} \bibinfo{year}{2013}\natexlab{}.
\newblock \showarticletitle{A multi-level typology of abstract visualization
  tasks}.
\newblock \bibinfo{journal}{\emph{IEEE transactions on visualization and
  computer graphics}} \bibinfo{volume}{19}, \bibinfo{number}{12}
  (\bibinfo{year}{2013}), \bibinfo{pages}{2376--2385}.
\newblock


\bibitem[Cabrera et~al\mbox{.}(2019)]%
        {cabrera2019fairvis}
\bibfield{author}{\bibinfo{person}{{\'A}ngel~Alexander Cabrera},
  \bibinfo{person}{Will Epperson}, \bibinfo{person}{Fred Hohman},
  \bibinfo{person}{Minsuk Kahng}, \bibinfo{person}{Jamie Morgenstern}, {and}
  \bibinfo{person}{Duen~Horng Chau}.} \bibinfo{year}{2019}\natexlab{}.
\newblock \showarticletitle{FairVis: Visual analytics for discovering
  intersectional bias in machine learning}. In \bibinfo{booktitle}{\emph{2019
  IEEE Conference on Visual Analytics Science and Technology (VAST)}}. IEEE,
  \bibinfo{pages}{46--56}.
\newblock


\bibitem[Cachel et~al\mbox{.}(2022)]%
        {cachel2022manirank}
\bibfield{author}{\bibinfo{person}{Kathleen Cachel}, \bibinfo{person}{Elke
  Rundensteiner}, {and} \bibinfo{person}{Lane Harrison}.}
  \bibinfo{year}{2022}\natexlab{}.
\newblock \showarticletitle{MANI-Rank: Multiple Attribute and Intersectional
  Group Fairness for Consensus Ranking}. In \bibinfo{booktitle}{\emph{2022 IEEE
  38th Intl. Conf. on Data Engineering (ICDE)}}. IEEE.
\newblock


\bibitem[Carenini and Loyd(2004)]%
        {carenini2004valuecharts}
\bibfield{author}{\bibinfo{person}{Giuseppe Carenini} {and}
  \bibinfo{person}{John Loyd}.} \bibinfo{year}{2004}\natexlab{}.
\newblock \showarticletitle{Valuecharts: analyzing linear models expressing
  preferences and evaluations}. In \bibinfo{booktitle}{\emph{Proceedings of the
  working conference on Advanced visual interfaces}}.
  \bibinfo{pages}{150--157}.
\newblock


\bibitem[Cheng et~al\mbox{.}(2021)]%
        {cheng2021soliciting}
\bibfield{author}{\bibinfo{person}{Hao-Fei Cheng}, \bibinfo{person}{Logan
  Stapleton}, \bibinfo{person}{Ruiqi Wang}, \bibinfo{person}{Paige Bullock},
  \bibinfo{person}{Alexandra Chouldechova}, \bibinfo{person}{Zhiwei
  Steven~Steven Wu}, {and} \bibinfo{person}{Haiyi Zhu}.}
  \bibinfo{year}{2021}\natexlab{}.
\newblock \showarticletitle{Soliciting stakeholders’ fairness notions in
  child maltreatment predictive systems}. In
  \bibinfo{booktitle}{\emph{Proceedings of the 2021 CHI Conference on Human
  Factors in Computing Systems}}. \bibinfo{pages}{1--17}.
\newblock


\bibitem[Constantin et~al\mbox{.}(2022)]%
        {constantin2022algorithmic}
\bibfield{author}{\bibinfo{person}{Rare{\c{s}} Constantin},
  \bibinfo{person}{Moritz D{\"u}ck}, \bibinfo{person}{Anton Alexandrov},
  \bibinfo{person}{Patrik Mato{\v{s}}evi{\'c}}, \bibinfo{person}{Daphna
  Keidar}, {and} \bibinfo{person}{Mennatallah El-Assady}.}
  \bibinfo{year}{2022}\natexlab{}.
\newblock \showarticletitle{How Do Algorithmic Fairness Metrics Align with
  Human Judgement? A Mixed-Initiative System for Contextualized Fairness
  Assessment}. In \bibinfo{booktitle}{\emph{2022 IEEE Workshop on TRust and
  EXpertise in Visual Analytics (TREX)}}. IEEE, \bibinfo{pages}{1--7}.
\newblock


\bibitem[Copeland(1951)]%
        {copeland1951reasonable}
\bibfield{author}{\bibinfo{person}{Arthur~H Copeland}.}
  \bibinfo{year}{1951}\natexlab{}.
\newblock \bibinfo{booktitle}{\emph{A reasonable social welfare function}}.
\newblock \bibinfo{type}{{T}echnical {R}eport}. \bibinfo{institution}{Mimeo,
  University of Michigan USA}.
\newblock


\bibitem[Crisan et~al\mbox{.}(2022)]%
        {crisan2022interactive}
\bibfield{author}{\bibinfo{person}{Anamaria Crisan}, \bibinfo{person}{Margaret
  Drouhard}, \bibinfo{person}{Jesse Vig}, {and} \bibinfo{person}{Nazneen
  Rajani}.} \bibinfo{year}{2022}\natexlab{}.
\newblock \showarticletitle{Interactive model cards: A human-centered approach
  to model documentation}. In \bibinfo{booktitle}{\emph{2022 ACM Conference on
  Fairness, Accountability, and Transparency}}. \bibinfo{pages}{427--439}.
\newblock


\bibitem[Deng et~al\mbox{.}(2022)]%
        {deng2022exploring}
\bibfield{author}{\bibinfo{person}{Wesley~Hanwen Deng}, \bibinfo{person}{Manish
  Nagireddy}, \bibinfo{person}{Michelle Seng~Ah Lee}, \bibinfo{person}{Jatinder
  Singh}, \bibinfo{person}{Zhiwei~Steven Wu}, \bibinfo{person}{Kenneth
  Holstein}, {and} \bibinfo{person}{Haiyi Zhu}.}
  \bibinfo{year}{2022}\natexlab{}.
\newblock \showarticletitle{Exploring how machine learning practitioners (try
  to) use fairness toolkits}. In \bibinfo{booktitle}{\emph{2022 ACM Conference
  on Fairness, Accountability, and Transparency}}. \bibinfo{pages}{473--484}.
\newblock


\bibitem[Dimara et~al\mbox{.}(2017)]%
        {dimara2017dcpairs}
\bibfield{author}{\bibinfo{person}{Evanthia Dimara}, \bibinfo{person}{Paola
  Valdivia}, {and} \bibinfo{person}{Christoph Kinkeldey}.}
  \bibinfo{year}{2017}\natexlab{}.
\newblock \showarticletitle{Dcpairs: A pairs plot based decision support
  system}. In \bibinfo{booktitle}{\emph{EuroVis-19th EG/VGTC Conference on
  Visualization}}.
\newblock


\bibitem[Gratzl et~al\mbox{.}(2013)]%
        {gratzl2013lineup}
\bibfield{author}{\bibinfo{person}{Samuel Gratzl}, \bibinfo{person}{Alexander
  Lex}, \bibinfo{person}{Nils Gehlenborg}, \bibinfo{person}{Hanspeter Pfister},
  {and} \bibinfo{person}{Marc Streit}.} \bibinfo{year}{2013}\natexlab{}.
\newblock \showarticletitle{Lineup: Visual analysis of multi-attribute
  rankings}.
\newblock \bibinfo{journal}{\emph{IEEE transactions on visualization and
  computer graphics}} \bibinfo{volume}{19}, \bibinfo{number}{12}
  (\bibinfo{year}{2013}), \bibinfo{pages}{2277--2286}.
\newblock


\bibitem[Hannan et~al\mbox{.}(2021)]%
        {hannan2021gets}
\bibfield{author}{\bibinfo{person}{Jacqueline Hannan},
  \bibinfo{person}{Huei-Yen~Winnie Chen}, {and} \bibinfo{person}{Kenneth
  Joseph}.} \bibinfo{year}{2021}\natexlab{}.
\newblock \showarticletitle{Who Gets What, According to Whom? An Analysis of
  Fairness Perceptions in Service Allocation}. In
  \bibinfo{booktitle}{\emph{Proceedings of the 2021 AAAI/ACM Conference on AI,
  Ethics, and Society}}. \bibinfo{pages}{555--565}.
\newblock


\bibitem[Hansen and Ombler(2008)]%
        {hansen2008new}
\bibfield{author}{\bibinfo{person}{Paul Hansen} {and} \bibinfo{person}{Franz
  Ombler}.} \bibinfo{year}{2008}\natexlab{}.
\newblock \showarticletitle{A new method for scoring additive multi-attribute
  value models using pairwise rankings of alternatives}.
\newblock \bibinfo{journal}{\emph{Journal of Multi-Criteria Decision Analysis}}
  \bibinfo{volume}{15}, \bibinfo{number}{3-4} (\bibinfo{year}{2008}),
  \bibinfo{pages}{87--107}.
\newblock


\bibitem[Harrison et~al\mbox{.}(2020)]%
        {harrison2020empirical}
\bibfield{author}{\bibinfo{person}{Galen Harrison}, \bibinfo{person}{Julia
  Hanson}, \bibinfo{person}{Christine Jacinto}, \bibinfo{person}{Julio
  Ramirez}, {and} \bibinfo{person}{Blase Ur}.} \bibinfo{year}{2020}\natexlab{}.
\newblock \showarticletitle{An empirical study on the perceived fairness of
  realistic, imperfect machine learning models}. In
  \bibinfo{booktitle}{\emph{Proceedings of the 2020 conference on fairness,
  accountability, and transparency}}. \bibinfo{pages}{392--402}.
\newblock


\bibitem[Hayez et~al\mbox{.}(2012)]%
        {hayez2012d}
\bibfield{author}{\bibinfo{person}{Quantin Hayez}, \bibinfo{person}{Yves
  De~Smet}, {and} \bibinfo{person}{Jimmy Bonney}.}
  \bibinfo{year}{2012}\natexlab{}.
\newblock \showarticletitle{D-Sight: a new decision making software to address
  multi-criteria problems}.
\newblock \bibinfo{journal}{\emph{International Journal of Decision Support
  System Technology (IJDSST)}} \bibinfo{volume}{4}, \bibinfo{number}{4}
  (\bibinfo{year}{2012}), \bibinfo{pages}{1--23}.
\newblock


\bibitem[Hindalong et~al\mbox{.}(2020)]%
        {hindalong2020towards}
\bibfield{author}{\bibinfo{person}{Emily Hindalong}, \bibinfo{person}{Jordon
  Johnson}, \bibinfo{person}{Giuseppe Carenini}, {and} \bibinfo{person}{Tamara
  Munzner}.} \bibinfo{year}{2020}\natexlab{}.
\newblock \showarticletitle{Towards Rigorously Designed Preference
  Visualizations for Group Decision Making}. In \bibinfo{booktitle}{\emph{2020
  IEEE Pacific Visualization Symposium (PacificVis)}}. IEEE,
  \bibinfo{pages}{181--190}.
\newblock


\bibitem[Hindalong et~al\mbox{.}(2022)]%
        {hindalong2022abstractions}
\bibfield{author}{\bibinfo{person}{Emily Hindalong}, \bibinfo{person}{Jordon
  Johnson}, \bibinfo{person}{Giuseppe Carenini}, {and} \bibinfo{person}{Tamara
  Munzner}.} \bibinfo{year}{2022}\natexlab{}.
\newblock \showarticletitle{Abstractions for Visualizing Preferences in Group
  Decisions}.
\newblock \bibinfo{journal}{\emph{Proceedings of the ACM on Human-Computer
  Interaction}} \bibinfo{volume}{6}, \bibinfo{number}{CSCW1}
  (\bibinfo{year}{2022}), \bibinfo{pages}{1--44}.
\newblock


\bibitem[Hong et~al\mbox{.}(2018)]%
        {hong2018collaborative}
\bibfield{author}{\bibinfo{person}{Sungsoo Hong}, \bibinfo{person}{Minhyang
  Suh}, \bibinfo{person}{Nathalie Henry~Riche}, \bibinfo{person}{Jooyoung Lee},
  \bibinfo{person}{Juho Kim}, {and} \bibinfo{person}{Mark Zachry}.}
  \bibinfo{year}{2018}\natexlab{}.
\newblock \showarticletitle{Collaborative dynamic queries: Supporting
  distributed small group decision-making}. In
  \bibinfo{booktitle}{\emph{Proceedings of the 2018 CHI Conference on Human
  Factors in Computing Systems}}. \bibinfo{pages}{1--12}.
\newblock


\bibitem[Johnson and Brun(2022)]%
        {johnson2022fairkit}
\bibfield{author}{\bibinfo{person}{Brittany Johnson} {and}
  \bibinfo{person}{Yuriy Brun}.} \bibinfo{year}{2022}\natexlab{}.
\newblock \showarticletitle{Fairkit-learn: a fairness evaluation and comparison
  toolkit}. In \bibinfo{booktitle}{\emph{Proceedings of the ACM/IEEE 44th
  International Conference on Software Engineering: Companion Proceedings}}.
  \bibinfo{pages}{70--74}.
\newblock


\bibitem[Kemeny(1959)]%
        {kemeny1959mathematics}
\bibfield{author}{\bibinfo{person}{John~G Kemeny}.}
  \bibinfo{year}{1959}\natexlab{}.
\newblock \showarticletitle{Mathematics without numbers}.
\newblock \bibinfo{journal}{\emph{Daedalus}} \bibinfo{volume}{88},
  \bibinfo{number}{4} (\bibinfo{year}{1959}), \bibinfo{pages}{577--591}.
\newblock


\bibitem[Kendall(1938)]%
        {kendall1938new}
\bibfield{author}{\bibinfo{person}{Maurice~G Kendall}.}
  \bibinfo{year}{1938}\natexlab{}.
\newblock \showarticletitle{A new measure of rank correlation}.
\newblock \bibinfo{journal}{\emph{Biometrika}} \bibinfo{volume}{30},
  \bibinfo{number}{1/2} (\bibinfo{year}{1938}), \bibinfo{pages}{81--93}.
\newblock


\bibitem[Kimmons(2012)]%
        {kimmons}
\bibfield{author}{\bibinfo{person}{Royce Kimmons}.}
  \bibinfo{year}{2012}\natexlab{}.
\newblock \bibinfo{title}{Exam scores}.
\newblock
\newblock
\urldef\tempurl%
\url{http://roycekimmons.com/tools/generated_data/exams}
\showURL{%
\tempurl}


\bibitem[Kuhlman and Rundensteiner(2020)]%
        {kuhlman2020rank}
\bibfield{author}{\bibinfo{person}{Caitlin Kuhlman} {and} \bibinfo{person}{Elke
  Rundensteiner}.} \bibinfo{year}{2020}\natexlab{}.
\newblock \showarticletitle{Rank aggregation algorithms for fair consensus}.
\newblock \bibinfo{journal}{\emph{Proceedings of the VLDB Endowment}}
  \bibinfo{volume}{13}, \bibinfo{number}{12} (\bibinfo{year}{2020}).
\newblock


\bibitem[Lam et~al\mbox{.}(2017)]%
        {lam2017bridging}
\bibfield{author}{\bibinfo{person}{Heidi Lam}, \bibinfo{person}{Melanie Tory},
  {and} \bibinfo{person}{Tamara Munzner}.} \bibinfo{year}{2017}\natexlab{}.
\newblock \showarticletitle{Bridging from goals to tasks with design study
  analysis reports}.
\newblock \bibinfo{journal}{\emph{IEEE trans. on visualization and computer
  graphics}} \bibinfo{volume}{24}, \bibinfo{number}{1} (\bibinfo{year}{2017}),
  \bibinfo{pages}{435--445}.
\newblock


\bibitem[Lee and Singh(2021)]%
        {lee2021landscape}
\bibfield{author}{\bibinfo{person}{Michelle Seng~Ah Lee} {and}
  \bibinfo{person}{Jat Singh}.} \bibinfo{year}{2021}\natexlab{}.
\newblock \showarticletitle{The landscape and gaps in open source fairness
  toolkits}. In \bibinfo{booktitle}{\emph{Proceedings of the 2021 CHI
  conference on human factors in computing systems}}. \bibinfo{pages}{1--13}.
\newblock


\bibitem[Liu et~al\mbox{.}(2018)]%
        {liu2018consensus}
\bibfield{author}{\bibinfo{person}{Weichen Liu}, \bibinfo{person}{Sijia Xiao},
  \bibinfo{person}{Jacob~T Browne}, \bibinfo{person}{Ming Yang}, {and}
  \bibinfo{person}{Steven~P Dow}.} \bibinfo{year}{2018}\natexlab{}.
\newblock \showarticletitle{ConsensUs: Supporting multi-criteria group
  decisions by visualizing points of disagreement}.
\newblock \bibinfo{journal}{\emph{ACM Transactions on Social Computing}}
  \bibinfo{volume}{1}, \bibinfo{number}{1} (\bibinfo{year}{2018}),
  \bibinfo{pages}{1--26}.
\newblock


\bibitem[Maguire et~al\mbox{.}(2012)]%
        {maguire2012taxonomy}
\bibfield{author}{\bibinfo{person}{Eamonn Maguire}, \bibinfo{person}{Philippe
  Rocca-Serra}, \bibinfo{person}{Susanna-Assunta Sansone}, \bibinfo{person}{Jim
  Davies}, {and} \bibinfo{person}{Min Chen}.} \bibinfo{year}{2012}\natexlab{}.
\newblock \showarticletitle{Taxonomy-based glyph design—with a case study on
  visualizing workflows of biological experiments}.
\newblock \bibinfo{journal}{\emph{IEEE Transactions on Visualization and
  Computer Graphics}} \bibinfo{volume}{18}, \bibinfo{number}{12}
  (\bibinfo{year}{2012}), \bibinfo{pages}{2603--2612}.
\newblock


\bibitem[Mashhadi et~al\mbox{.}(2022)]%
        {mashhadi2022case}
\bibfield{author}{\bibinfo{person}{Afra Mashhadi}, \bibinfo{person}{Annuska
  Zolyomi}, {and} \bibinfo{person}{Jay Quedado}.}
  \bibinfo{year}{2022}\natexlab{}.
\newblock \showarticletitle{A Case Study of Integrating Fairness Visualization
  Tools in Machine Learning Education}. In \bibinfo{booktitle}{\emph{CHI
  Conference on Human Factors in Computing Systems Extended Abstracts}}.
  \bibinfo{pages}{1--7}.
\newblock


\bibitem[Mitchell et~al\mbox{.}(2019)]%
        {mitchell2019model}
\bibfield{author}{\bibinfo{person}{Margaret Mitchell}, \bibinfo{person}{Simone
  Wu}, \bibinfo{person}{Andrew Zaldivar}, \bibinfo{person}{Parker Barnes},
  \bibinfo{person}{Lucy Vasserman}, \bibinfo{person}{Ben Hutchinson},
  \bibinfo{person}{Elena Spitzer}, \bibinfo{person}{Inioluwa~Deborah Raji},
  {and} \bibinfo{person}{Timnit Gebru}.} \bibinfo{year}{2019}\natexlab{}.
\newblock \showarticletitle{Model cards for model reporting}. In
  \bibinfo{booktitle}{\emph{Proceedings of the conference on fairness,
  accountability, and transparency}}. \bibinfo{pages}{220--229}.
\newblock


\bibitem[Mulligan et~al\mbox{.}(2019)]%
        {mulligan2019thing}
\bibfield{author}{\bibinfo{person}{Deirdre~K Mulligan},
  \bibinfo{person}{Joshua~A Kroll}, \bibinfo{person}{Nitin Kohli}, {and}
  \bibinfo{person}{Richmond~Y Wong}.} \bibinfo{year}{2019}\natexlab{}.
\newblock \showarticletitle{This thing called fairness: Disciplinary confusion
  realizing a value in technology}.
\newblock \bibinfo{journal}{\emph{Proceedings of the ACM on Human-Computer
  Interaction}} \bibinfo{volume}{3}, \bibinfo{number}{CSCW}
  (\bibinfo{year}{2019}), \bibinfo{pages}{1--36}.
\newblock


\bibitem[Munechika et~al\mbox{.}(2022)]%
        {munechika2022visual}
\bibfield{author}{\bibinfo{person}{David Munechika}, \bibinfo{person}{Zijie~J
  Wang}, \bibinfo{person}{Jack Reidy}, \bibinfo{person}{Josh Rubin},
  \bibinfo{person}{Krishna Gade}, \bibinfo{person}{Krishnaram Kenthapadi},
  {and} \bibinfo{person}{Duen~Horng Chau}.} \bibinfo{year}{2022}\natexlab{}.
\newblock \showarticletitle{Visual Auditor: Interactive Visualization for
  Detection and Summarization of Model Biases}. In
  \bibinfo{booktitle}{\emph{2022 IEEE Visualization and Visual Analytics
  (VIS)}}. IEEE, \bibinfo{pages}{45--49}.
\newblock


\bibitem[Munzner(2009)]%
        {munzner2009nested}
\bibfield{author}{\bibinfo{person}{Tamara Munzner}.}
  \bibinfo{year}{2009}\natexlab{}.
\newblock \showarticletitle{A nested model for visualization design and
  validation}.
\newblock \bibinfo{journal}{\emph{IEEE transactions on visualization and
  computer graphics}} \bibinfo{volume}{15}, \bibinfo{number}{6}
  (\bibinfo{year}{2009}), \bibinfo{pages}{921--928}.
\newblock


\bibitem[Mustajoki and H{\"a}m{\"a}l{\"a}inen(2000)]%
        {mustajoki2000web}
\bibfield{author}{\bibinfo{person}{Jyri Mustajoki} {and}
  \bibinfo{person}{Raimo~P H{\"a}m{\"a}l{\"a}inen}.}
  \bibinfo{year}{2000}\natexlab{}.
\newblock \showarticletitle{Web-HIPRE: Global decision support by value tree
  and AHP analysis}.
\newblock \bibinfo{journal}{\emph{INFOR: Information Systems and Operational
  Research}} \bibinfo{volume}{38}, \bibinfo{number}{3} (\bibinfo{year}{2000}),
  \bibinfo{pages}{208--220}.
\newblock


\bibitem[Nobre et~al\mbox{.}(2020)]%
        {nobre2020evaluating}
\bibfield{author}{\bibinfo{person}{Carolina Nobre}, \bibinfo{person}{Dylan
  Wootton}, \bibinfo{person}{Lane Harrison}, {and} \bibinfo{person}{Alexander
  Lex}.} \bibinfo{year}{2020}\natexlab{}.
\newblock \showarticletitle{Evaluating multivariate network visualization
  techniques using a validated design and crowdsourcing approach}. In
  \bibinfo{booktitle}{\emph{Proceedings of the 2020 CHI conference on human
  factors in computing systems}}. \bibinfo{pages}{1--12}.
\newblock


\bibitem[Pham and Huang(2016)]%
        {pham2016qstack}
\bibfield{author}{\bibinfo{person}{Phi~Giang Pham} {and}
  \bibinfo{person}{Mao~Lin Huang}.} \bibinfo{year}{2016}\natexlab{}.
\newblock \showarticletitle{Qstack: Multi-tag Visual Rankings.}
\newblock \bibinfo{journal}{\emph{J. Softw.}} \bibinfo{volume}{11},
  \bibinfo{number}{7} (\bibinfo{year}{2016}), \bibinfo{pages}{695--703}.
\newblock


\bibitem[Richardson et~al\mbox{.}(2021)]%
        {richardson2021towards}
\bibfield{author}{\bibinfo{person}{Brianna Richardson}, \bibinfo{person}{Jean
  Garcia-Gathright}, \bibinfo{person}{Samuel~F Way}, \bibinfo{person}{Jennifer
  Thom}, {and} \bibinfo{person}{Henriette Cramer}.}
  \bibinfo{year}{2021}\natexlab{}.
\newblock \showarticletitle{Towards fairness in practice: A
  practitioner-oriented rubric for evaluating Fair ML Toolkits}. In
  \bibinfo{booktitle}{\emph{Proceedings of the 2021 CHI Conference on Human
  Factors in Computing Systems}}. \bibinfo{pages}{1--13}.
\newblock


\bibitem[Saleiro et~al\mbox{.}(2018)]%
        {saleiro2018aequitas}
\bibfield{author}{\bibinfo{person}{Pedro Saleiro}, \bibinfo{person}{Benedict
  Kuester}, \bibinfo{person}{Loren Hinkson}, \bibinfo{person}{Jesse London},
  \bibinfo{person}{Abby Stevens}, \bibinfo{person}{Ari Anisfeld},
  \bibinfo{person}{Kit~T Rodolfa}, {and} \bibinfo{person}{Rayid Ghani}.}
  \bibinfo{year}{2018}\natexlab{}.
\newblock \showarticletitle{Aequitas: A bias and fairness audit toolkit}.
\newblock \bibinfo{journal}{\emph{arXiv preprint arXiv:1811.05577}}
  (\bibinfo{year}{2018}).
\newblock


\bibitem[Saxena et~al\mbox{.}(2019)]%
        {saxena2019fairness}
\bibfield{author}{\bibinfo{person}{Nripsuta~Ani Saxena}, \bibinfo{person}{Karen
  Huang}, \bibinfo{person}{Evan DeFilippis}, \bibinfo{person}{Goran Radanovic},
  \bibinfo{person}{David~C Parkes}, {and} \bibinfo{person}{Yang Liu}.}
  \bibinfo{year}{2019}\natexlab{}.
\newblock \showarticletitle{How do fairness definitions fare? Examining public
  attitudes towards algorithmic definitions of fairness}. In
  \bibinfo{booktitle}{\emph{Proceedings of the 2019 AAAI/ACM Conference on AI,
  Ethics, and Society}}. \bibinfo{pages}{99--106}.
\newblock


\bibitem[Schulze(2018)]%
        {schulze2018schulze}
\bibfield{author}{\bibinfo{person}{Markus Schulze}.}
  \bibinfo{year}{2018}\natexlab{}.
\newblock \showarticletitle{The Schulze method of voting}.
\newblock \bibinfo{journal}{\emph{arXiv preprint arXiv:1804.02973}}
  (\bibinfo{year}{2018}).
\newblock


\bibitem[Shah(2014)]%
        {shah2014collaborative}
\bibfield{author}{\bibinfo{person}{Chirag Shah}.}
  \bibinfo{year}{2014}\natexlab{}.
\newblock \showarticletitle{Collaborative information seeking}.
\newblock \bibinfo{journal}{\emph{Journal of the Association for Information
  Science and Technology}} \bibinfo{volume}{65}, \bibinfo{number}{2}
  (\bibinfo{year}{2014}), \bibinfo{pages}{215--236}.
\newblock


\bibitem[Shrestha et~al\mbox{.}(2022)]%
        {shrestha2022fairfuse}
\bibfield{author}{\bibinfo{person}{Hilson Shrestha}, \bibinfo{person}{Kathleen
  Cachel}, \bibinfo{person}{Mallak Alkhathlan}, \bibinfo{person}{Elke
  Rundensteiner}, {and} \bibinfo{person}{Lane Harrison}.}
  \bibinfo{year}{2022}\natexlab{}.
\newblock \showarticletitle{FairFuse: Interactive Visual Support for Fair
  Consensus Ranking}. In \bibinfo{booktitle}{\emph{2022 IEEE Visualization and
  Visual Analytics (VIS)}}. IEEE, \bibinfo{pages}{65--69}.
\newblock


\bibitem[Srivastava et~al\mbox{.}(2019)]%
        {srivastava2019mathematical}
\bibfield{author}{\bibinfo{person}{Megha Srivastava}, \bibinfo{person}{Hoda
  Heidari}, {and} \bibinfo{person}{Andreas Krause}.}
  \bibinfo{year}{2019}\natexlab{}.
\newblock \showarticletitle{Mathematical notions vs. human perception of
  fairness: A descriptive approach to fairness for machine learning}. In
  \bibinfo{booktitle}{\emph{Proceedings of the 25th ACM SIGKDD international
  conference on knowledge discovery \& data mining}}.
  \bibinfo{pages}{2459--2468}.
\newblock


\bibitem[Tramer et~al\mbox{.}(2017)]%
        {tramer2017fairtest}
\bibfield{author}{\bibinfo{person}{Florian Tramer}, \bibinfo{person}{Vaggelis
  Atlidakis}, \bibinfo{person}{Roxana Geambasu}, \bibinfo{person}{Daniel Hsu},
  \bibinfo{person}{Jean-Pierre Hubaux}, \bibinfo{person}{Mathias Humbert},
  \bibinfo{person}{Ari Juels}, {and} \bibinfo{person}{Huang Lin}.}
  \bibinfo{year}{2017}\natexlab{}.
\newblock \showarticletitle{Fairtest: Discovering unwarranted associations in
  data-driven applications}. In \bibinfo{booktitle}{\emph{2017 IEEE European
  Symposium on Security and Privacy (EuroS\&P)}}. IEEE,
  \bibinfo{pages}{401--416}.
\newblock


\bibitem[Van~Berkel et~al\mbox{.}(2021)]%
        {van2021effect}
\bibfield{author}{\bibinfo{person}{Niels Van~Berkel}, \bibinfo{person}{Jorge
  Goncalves}, \bibinfo{person}{Daniel Russo}, \bibinfo{person}{Simo Hosio},
  {and} \bibinfo{person}{Mikael~B Skov}.} \bibinfo{year}{2021}\natexlab{}.
\newblock \showarticletitle{Effect of information presentation on fairness
  perceptions of machine learning predictors}. In
  \bibinfo{booktitle}{\emph{Proceedings of the 2021 CHI Conference on Human
  Factors in Computing Systems}}. \bibinfo{pages}{1--13}.
\newblock


\bibitem[Verma and Rubin(2018)]%
        {verma2018fairness}
\bibfield{author}{\bibinfo{person}{Sahil Verma} {and} \bibinfo{person}{Julia
  Rubin}.} \bibinfo{year}{2018}\natexlab{}.
\newblock \showarticletitle{Fairness definitions explained}. In
  \bibinfo{booktitle}{\emph{2018 ieee/acm international workshop on software
  fairness (fairware)}}. IEEE, \bibinfo{pages}{1--7}.
\newblock


\bibitem[Wall et~al\mbox{.}(2017)]%
        {wall2017podium}
\bibfield{author}{\bibinfo{person}{Emily Wall}, \bibinfo{person}{Subhajit Das},
  \bibinfo{person}{Ravish Chawla}, \bibinfo{person}{Bharath Kalidindi},
  \bibinfo{person}{Eli~T Brown}, {and} \bibinfo{person}{Alex Endert}.}
  \bibinfo{year}{2017}\natexlab{}.
\newblock \showarticletitle{Podium: Ranking data using mixed-initiative visual
  analytics}.
\newblock \bibinfo{journal}{\emph{IEEE transactions on visualization and
  computer graphics}} \bibinfo{volume}{24}, \bibinfo{number}{1}
  (\bibinfo{year}{2017}), \bibinfo{pages}{288--297}.
\newblock


\bibitem[Weng et~al\mbox{.}(2018a)]%
        {weng2018srvis}
\bibfield{author}{\bibinfo{person}{Di Weng}, \bibinfo{person}{Ran Chen},
  \bibinfo{person}{Zikun Deng}, \bibinfo{person}{Feiran Wu},
  \bibinfo{person}{Jingmin Chen}, {and} \bibinfo{person}{Yingcai Wu}.}
  \bibinfo{year}{2018}\natexlab{a}.
\newblock \showarticletitle{Srvis: Towards better spatial integration in
  ranking visualization}.
\newblock \bibinfo{journal}{\emph{IEEE transactions on visualization and
  computer graphics}} \bibinfo{volume}{25}, \bibinfo{number}{1}
  (\bibinfo{year}{2018}), \bibinfo{pages}{459--469}.
\newblock


\bibitem[Weng et~al\mbox{.}(2018b)]%
        {weng2018homefinder}
\bibfield{author}{\bibinfo{person}{Di Weng}, \bibinfo{person}{Heming Zhu},
  \bibinfo{person}{Jie Bao}, \bibinfo{person}{Yu Zheng}, {and}
  \bibinfo{person}{Yingcai Wu}.} \bibinfo{year}{2018}\natexlab{b}.
\newblock \showarticletitle{Homefinder revisited: Finding ideal homes with
  reachability-centric multi-criteria decision making}. In
  \bibinfo{booktitle}{\emph{Proceedings of the 2018 CHI Conference on Human
  Factors in Computing Systems}}. \bibinfo{pages}{1--12}.
\newblock


\bibitem[Wexler et~al\mbox{.}(2019)]%
        {wexler2019if}
\bibfield{author}{\bibinfo{person}{James Wexler}, \bibinfo{person}{Mahima
  Pushkarna}, \bibinfo{person}{Tolga Bolukbasi}, \bibinfo{person}{Martin
  Wattenberg}, \bibinfo{person}{Fernanda Vi{\'e}gas}, {and}
  \bibinfo{person}{Jimbo Wilson}.} \bibinfo{year}{2019}\natexlab{}.
\newblock \showarticletitle{The what-if tool: Interactive probing of machine
  learning models}.
\newblock \bibinfo{journal}{\emph{IEEE transactions on visualization and
  computer graphics}} \bibinfo{volume}{26}, \bibinfo{number}{1}
  (\bibinfo{year}{2019}), \bibinfo{pages}{56--65}.
\newblock


\bibitem[Xie et~al\mbox{.}(2021)]%
        {xie2021fairrankvis}
\bibfield{author}{\bibinfo{person}{Tiankai Xie}, \bibinfo{person}{Yuxin Ma},
  \bibinfo{person}{Jian Kang}, \bibinfo{person}{Hanghang Tong}, {and}
  \bibinfo{person}{Ross Maciejewski}.} \bibinfo{year}{2021}\natexlab{}.
\newblock \showarticletitle{FairRankVis: A Visual Analytics Framework for
  Exploring Algorithmic Fairness in Graph Mining Models}.
\newblock \bibinfo{journal}{\emph{IEEE Transactions on Visualization and
  Computer Graphics}} \bibinfo{volume}{28}, \bibinfo{number}{1}
  (\bibinfo{year}{2021}), \bibinfo{pages}{368--377}.
\newblock


\bibitem[Yang et~al\mbox{.}(2018)]%
        {yang2018nutritional}
\bibfield{author}{\bibinfo{person}{Ke Yang}, \bibinfo{person}{Julia
  Stoyanovich}, \bibinfo{person}{Abolfazl Asudeh}, \bibinfo{person}{Bill Howe},
  \bibinfo{person}{HV Jagadish}, {and} \bibinfo{person}{Gerome Miklau}.}
  \bibinfo{year}{2018}\natexlab{}.
\newblock \showarticletitle{A nutritional label for rankings}. In
  \bibinfo{booktitle}{\emph{Proceedings of the 2018 international conference on
  management of data}}. \bibinfo{pages}{1773--1776}.
\newblock


\end{thebibliography}


\clearpage
\appendix

\section{Demographics and Results Summary}

\begin{figure}[H]
    \includegraphics[width=\linewidth]{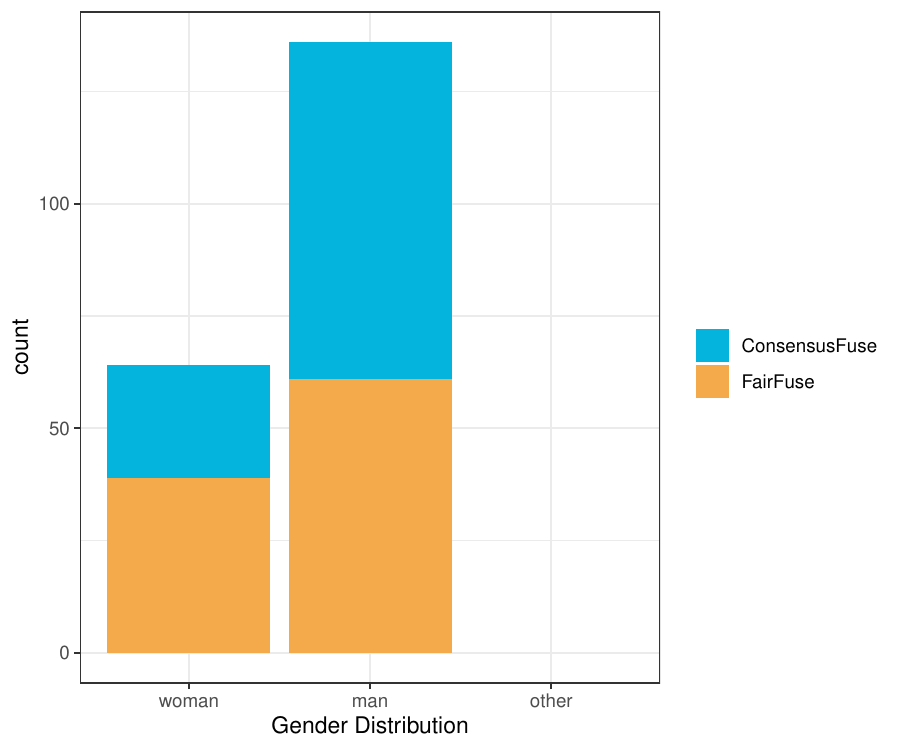}
    \caption{Gender Distribution}
    \label{fig:gender-distribution}
\end{figure}

\begin{figure}[H]
    \includegraphics[width=\linewidth]{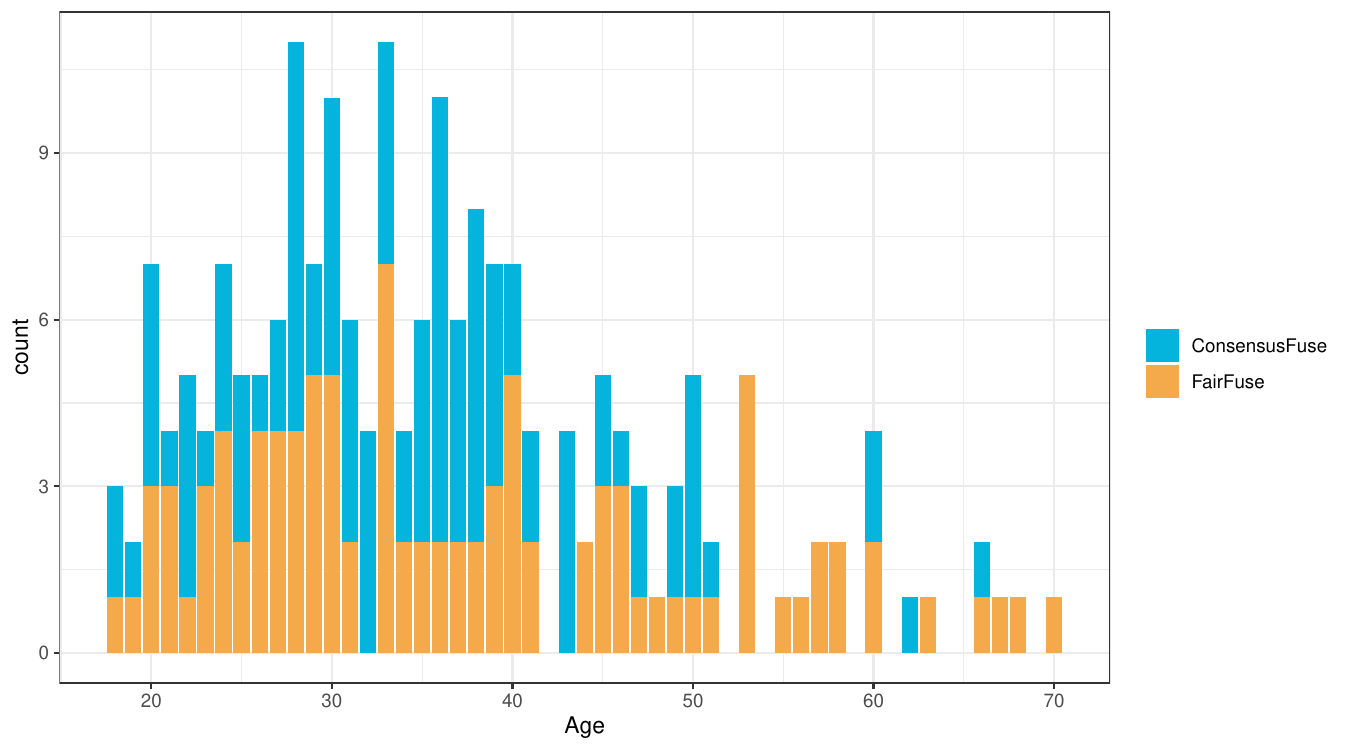}
    \caption{Age Distribution}
    \label{fig:age-distribution}

\end{figure}

\begin{figure}[H]
    \includegraphics[width=\linewidth]{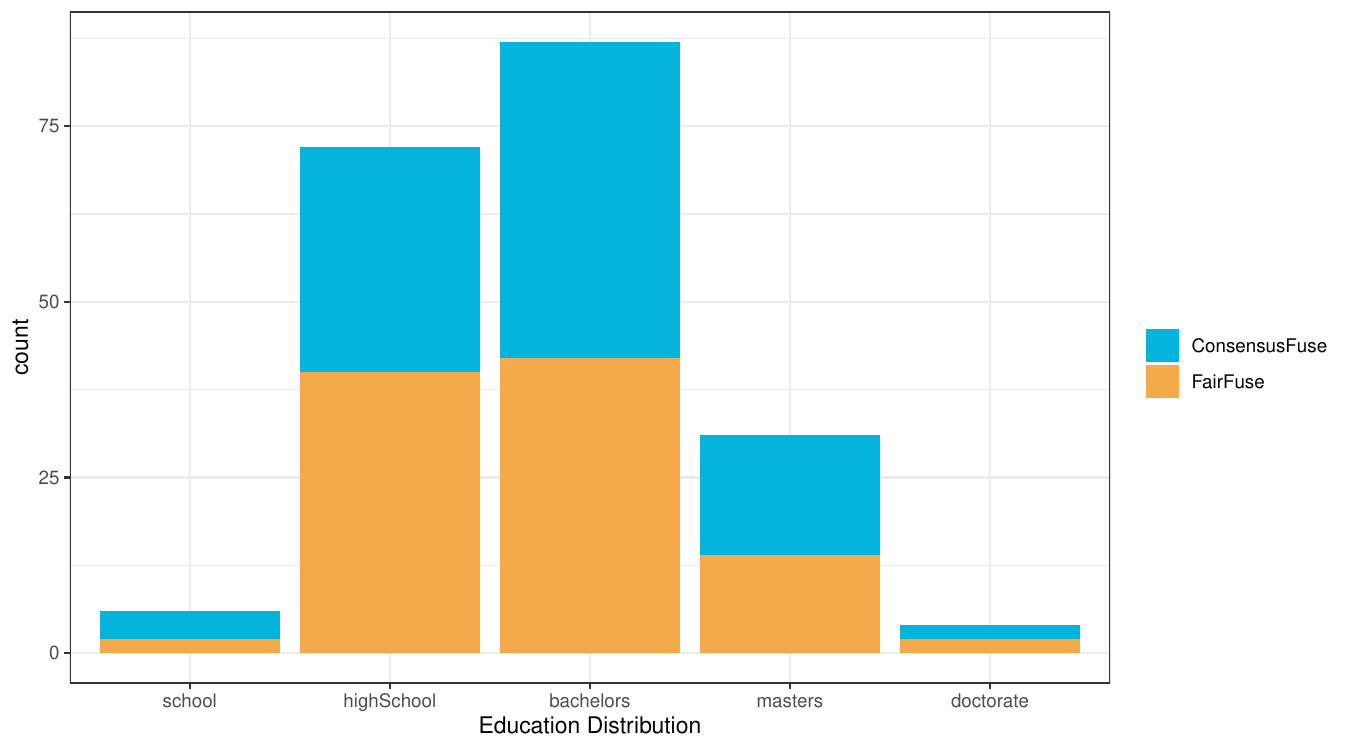}
    \caption{Education Level Distribution}
    \label{fig:education-distribution}
\end{figure}

\begin{figure}[H]
    \includegraphics[width=\linewidth]{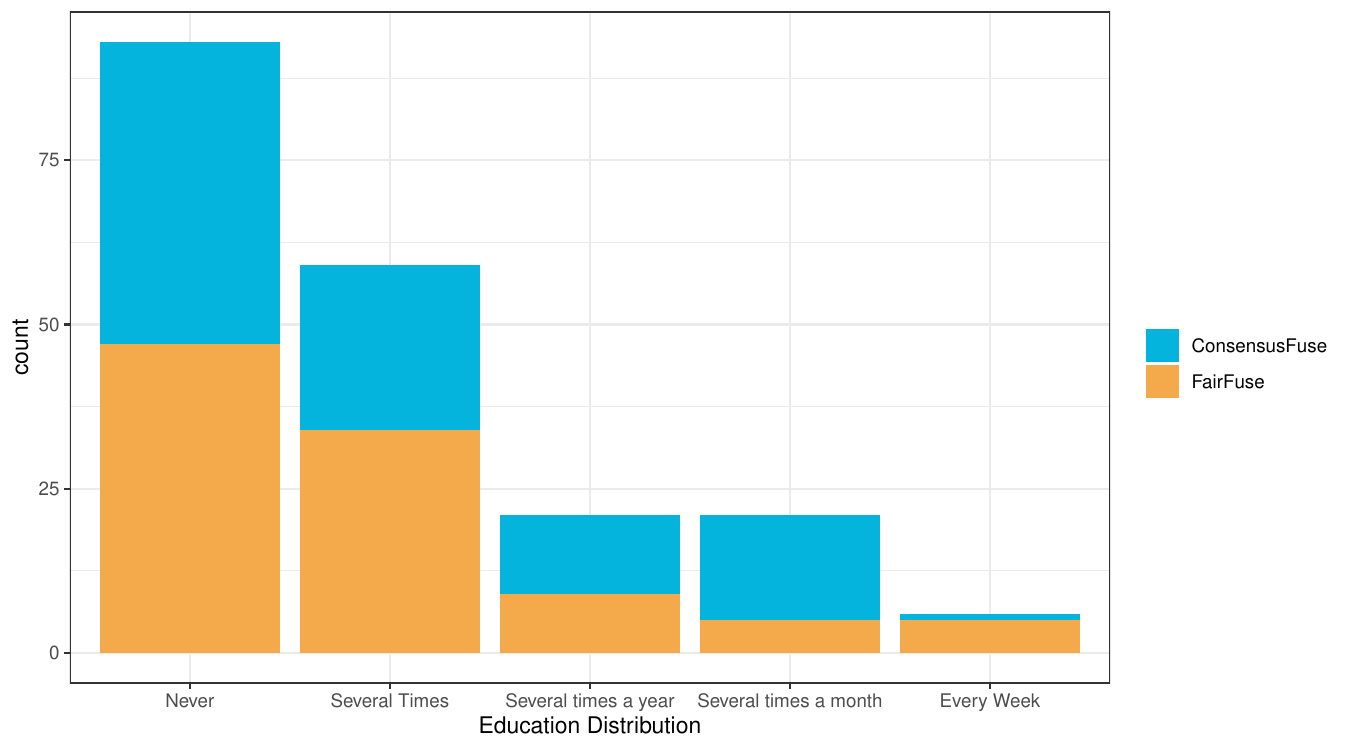}
    \caption{Visualization Experience Distribution}
    \label{fig:vis-exp-distribution}
\end{figure}


\begin{table*}[!h]
  \caption{Summary of the results by tasks }
  \label{tab:results-questions-participants}
  \centering
  \begin{tabular}{cp{.19\textwidth}lll}
  \toprule
  & \bfseries Task & \bfseries FairFuse & \bfseries ConsensusFuse & \bfseries Result \\
  \midrule
    \bfseries T1 & Locating protected attribute & $0.97\sim[0.91,0.99]$ & $1\sim[1,1]$ & $W=5150$ $p=0.0827$ $d=-0.247\sim[-0.527,0.0325]$ \\
    \bfseries T2 & Identifying Advantaged Group(s) & $0.84\sim[0.74,0.89]$ & $0.23\sim[0.15,0.31]$ & $W=1950$ $p=6.42e-18$ $d=1.54\sim[1.22,1.86]$ \\
    & a) Correctly identify race 1 & $0.94\sim[0.86,0.97]$ & $0.29\sim[0.19,0.38]$ & $W=1750$ $p=4.49e-21$ $d=1.79\sim[1.46,2.12]$\\
    & b) Correctly identify race 2 & $0.88\sim[0.79,0.93]$ & $0.9\sim[0.82,0.94]$ & $W=5100$ $p=0.654$ $d=-0.0636\sim[-0.343,0.215]$ \\
    \bfseries T3 & Visualization Use & $0.92\sim[0.82,0.95]$ &$0.92\sim[0.85,0.96]$ & $W=5000$ $p=1$ $d=0\sim[-0.279,0.279]$\\
    \bfseries T4 & Identifying Attribute-level Unfairness & $0.83\sim[0.73,0.88]$ & $0.4\sim[0.29,0.49]$ & $W=2850$ $p=4.62e-10$ $d=0.98\sim[0.685,1.28$ \\
    \bfseries T5 & Identifying Group-level Unfairness & $0.97\sim[0.91,0.99]$ & $0.81\sim[0.71,0.87]$ & $W=4200$ $ p=0.000313$ $d=0.526\sim[0.243,0.81]$\\
    \bfseries T6 & Utilizing PCP Position Comparison & $0.84\sim[0.74,0.89]$ & $0.84~[0.75,0.89]$ & $W=5000$ $p=1$ $d=0\sim[-0.279,0.279]$\\
    \bfseries T7 & Interpreting PCP Gradient & $0.71\sim[0.6,0.78]$ & $0.74\sim[0.64,0.81]$ & $W=5150$ $p=0.637$ $d=-0.0669\sim[-0.346,0.212]$\\
    \bfseries T8 & Using Consensus Generation Procedure & $0.65\sim[0.54,0.73]$ & $0.45\sim[0.34,0.53]$ & $W=4000$ $p=0.00459$ $d=0.408\sim[0.127,0.69]$\\
    \bfseries T9 & Using Fair Consensus Generation Procedure \\
    & a) ARP & $0.15\sim[0.12,0.18]$ & $0.31\sim[0.29,0.33]$ & $W=8270.5$ $p=7.77e-16$ $d=-1.39\sim[-1.71,-1.08]$\\
    & b) PD Loss & $0.18\sim[0.17,0.18]$ & $0.17\sim[0.16,0.19]$ & $W=2987.5$ $p=7.52e-07$ $d=0.0935\sim[-0.186,0.373]$ \\
    & c) Interactions Count & $12.2\sim[10.1,15.08]$ & $18.76\sim[15.05,24.85]$ & $W=5638$ $p=0.119$ $d=-0.346\sim[-0.627,-0.0648]$\\

  \bottomrule
  \end{tabular}
\end{table*}

\end{document}